\documentclass[a4paper,12pt]{report}
\usepackage[margin=1in]{geometry}
\usepackage{amsmath,amssymb}
\usepackage{graphicx}
\usepackage{cite}
\usepackage{amsfonts}
\usepackage{float}
\usepackage{graphicx}
\usepackage{epstopdf}
\usepackage{hyperref}
\usepackage{setspace}
\begin{document}


\begin{titlepage}
    \begin{center}
        \vspace*{1cm}
        
        \textbf{\LARGE{Transverse spin and transverse momentum in structured optical fields}}
        

        \vspace{7.5cm}
        By\\
                \textbf{Sudipta Saha$^1$, Nirmalya Ghosh$^1$, and Subhasish Dutta Gupta$^{2,*}$}\\ 
        \vspace{0.5cm}
       {$^1$Department of Physical Sciences, Indian Institute of Science Education and Research Kolkata, Mohanpur-741 246, India}
       
        {$^2$School of Physics, University of Hyderabad, Hyderabad 500046, India\\ 
        	\vspace{0.5cm}
        	 $^*$Corresponding author: sdghyderabad@gmail.com}

      \vfill
        
 Submitted to Wiley-VCH Encyclopedia of Applied Physics, \url{http://dx.doi.org/10.1002/3527600434}

        \vspace{1.8cm}
        
        

    \end{center}
\end{titlepage}

\thispagestyle{empty}

\newpage
 \begin{center}
\large{\textbf{\underline{Abstract}}}
\end{center}
\vspace*{1.5cm}
It has been recently recognized that in addition to the conventional longitudinal angular momentum, structured (inhomogeneous) optical fields exhibit helicity-independent transverse spin angular momentum (SAM) and an unusual spin (circular polarization)-dependent transverse momentum, the so-called Belinfante’s spin momentum. Such highly nontrivial structure of the momentum and the spin densities in the structured optical fields (e.g., evanescent fields) has led to a number of fundamentally interesting and intricate phenomena, e.g., the quantum spin Hall effect of light and the optical spin-momentum locking in surface optical modes similar to that observed for electrons in topological insulators. In this review, we introduce the basic concepts and look into the genesis of transverse SAM and transverse spin momentum in structured light. We then discuss few illustrative examples of micro and nano optical systems where these illusive entities can be observed. The studied systems include planar and spherical micro and nano structures. We also investigate the ways and means of enhancing the elusive extraordinary spin. In particular, we show that  dispersion management leading to avoided crossing along with perfect absorption mediated by recently discovered coherent perfect absorption can positively influence the resonant enhancement of the transverse spin and spin momentum. The role of mode mixing and interference of neighboring transverse electric and transverse magnetic scattering modes of diverse micro and nano optical systems are illustrated with the selected examples. The results demonstrate possibilities for the enhancement of not only the magnitudes but also the spatial extent of transverse SAM and the transverse momentum components, which opens up interesting avenues for experimental detection of these illusive fundamental entities and may enhance the ensuing spin-based photonic applications.


\newpage                                          

\tableofcontents


\chapter*{Introduction}
\addcontentsline{toc}{chapter}{Introduction}
It is well known that light wave carries both linear and angular momentum (AM) \cite{allenbook,sdgbook,bliokhphyrep}. The plane wave or paraxial Gaussian beam typically carries longitudinal momentum and longitudinal spin AM (SAM, degree of circular polarization or helicity $\sigma=\pm 1$). This definition of SAM can be generalized to include elliptically polarized light characterized by $-1\leq \sigma \leq +1$. In addition to SAM, higher order Gaussian beams (such as the Laguerre-Gaussian modes) can also carry orbital angular momentum (OAM), which can have both intrinsic and extrinsic contributions. The optical momentum and the two kinds of the AM of light play crucial roles in various light-matter interactions  \cite{brevik1979,cohen1998atom,chu1998,pfeifer2007,ashkin2000,phillips1998nobel,grier2003revolution}, and the presence of these are manifested as radiation-pressure force and torque (respectively) experienced by probe particles or atoms placed inside the field \cite{ashkin1983,chaumet2000,berry2009}. Moreover, on conceptual grounds, coupling and inter-conversion between the spin and orbital AM degrees of freedom of light is expected under certain circumstances. This so-called spin orbit interaction (SOI) of light has been observed recently in various optical interactions leading to a number of interesting and intricate polarization effects \cite{natphot2015,bliokh2015,zayats2014}. Most of these SOI effects are typically associated with inter-conversion between the longitudinal (along the propagation direction specified by the wave vector $\mathbf{k}$) SAM and the longitudinal OAM of light \cite{bliokhphyrep}, recently it has been discovered that in addition to the conventional longitudinal AM, structured (inhomogeneous) optical fields can exhibit an unusual transverse (perpendicular to the propagation direction or $\mathbf{k}$) SAM, which is independent of the helicity \cite{bliokh2014,aiello2015,bekshaev2015,bauer2016,zayats2014,measuring2015,natphys_2016}. Moreover, such inhomogeneous fields also demonstrate helicity-dependent transverse momentum, the so-called Belinfante’s spin momentum \cite{belinfante1940}. Such highly nontrivial structure of the momentum and spin densities in the structured optical fields (e.g., evanescent fields) has led to a number of fundamentally interesting and intricate phenomena, e.g., the quantum spin Hall Effect of light and the optical spin-momentum locking in surface optical modes\cite{zubinoptica,zubinapl}. The spin-directional locking in the excitation of surface optical modes (e.g., for surface plasmon-polaritons at dielectric-metal interfaces) has been experimentally observed and explained using the helicity-independent transverse spin and helicity-dependent transverse momentum of evanescent waves. This spin-momentum locking phenomenon in surface optical modes has been identified as the quantum spin Hall effect of light and analogies have been drawn  with the corresponding surface states of topological insulators for electrons \cite{bliokh2015}.
\par	
The main dynamical properties of any optical field are the time-averaged densities of energy, momentum, and SAM, which are conventionally defined using the electric and the magnetic fields of light \cite{bornwolf,sdgbook,allenbook,bliokhphyrep}. The momentum is usually the canonical (orbital) momentum, which is responsible for the radiation pressure and is accordingly physically observable. The transverse spin-momentum, introduced by Belinfante within field theory \cite{belinfante1940}, on the other hand, is known to be a virtual quantity (does not transfer energy or exert optical pressure on dipolar particles), which only generates the spin AM. As subsequently discussed, the Poynting vector, by its definition, contains contributions of both the orbital and the spin momentum, the former being the directly observable momentum of light \cite{berry2009,berry2015}. Note that for ideal plane waves (having purely transverse electric and magnetic field components), the Poynting vector has only longitudinal component (along the propagation direction) solely due to the orbital momentum. For non-plane waves (structured or inhomogeneous fields), on the other hand, finite longitudinal components of the electric and the magnetic fields are the sources of the transverse spin momentum.  Moreover, the transverse SAM appears when the longitudinal field component is shifted in phase with respect to the transverse field components. Despite this relatively candid origin of the transverse SAM and transverse spin momentum in structured optical fields, until recently these unusual angular momentum entities have not been recognized or experimentally observed. As noted above, recent observation of these entities in the evanescent fields have revealed a plethora of fundamentally interesting spin optical effects and opened up a new paradigm of spin based photonic applications. The purpose of this review is to introduce the basic concepts and explore the genesis of transverse SAM and transverse spin momentum in structured light with illustrative examples of their manifestations in diverse micro and nano optical systems and discuss their potential spin-controlled photonic applications. The review also explores various possible means of enhancing the elusive transverse spin and momentum.
\par
The paper is organized as follows. In Chapter \ref{chapter1}, we introduce the basic concepts and investigate the origin of transverse SAM and transverse spin momentum in structured light. We then discuss the role of dispersion management, avoided crossing and coherent perfect absorption on these entities. In Chapter \ref{chapter2}, we provide illustrative examples of observation of transverse SAM and transverse spin momentum in planar structures, namely, in Sarid geometry and in gap plasmons. Transverse SAM and momentum in scattering of  plane waves from spherical particles are discussed in Chapter \ref{chapter3}. The influence of the localized plasmon resonances in metal nanoparticles and the higher order transverse electric (TE) and transverse magnetic (TM) scattering modes of dielectric microspheres on the transverse spin and momentum components are illustrated with selected examples. The influence of mode mixing in layered spherical scatterers and the effect of the focused radially and azimuthally polarized vector beams on the spin and the momentum components of the scattered light are also discussed. This is followed by a brief account on the experimental means for detecting these illusive entities in nano-plasmonic and other micro and nano optical systems 
(Chapter \ref{chapter4}). The paper concludes with an outlook of the potential spin controlled photonic applications of transverse spin and momentum of structured light. 
\chapter{Genesis of transverse spin and transverse momentum in structured light \label{chapter1}}
This Chapter is devoted to a basic understanding of the optical momentum and the origin of transverse spin with structured light in typical scenarios involving total internal reflection, surface or guided modes (in particular, surface plasmons) and paraxial beams. Recall that an `unstructured' light like a transverse plane wave can never have a SAM or spin momentum component perpendicular to the direction of propagation. As mentioned in the Introduction, the generic feature common to all the diverse scenarios mentioned above is essentially the `structure' of the optical field. A deeper look reveals that the essential structure owes its origin to the longitudinal component of the electric or the magnetic field which is phase shifted from the corresponding transverse component resulting in a polarization-type ellipse in the \emph{plane of incidence} \cite{bliokh2015}. Mathematically, the emergence of the longitudinal component of the field for inhomogeneous waves is a direct consequence of the Maxwell's divergence equation. The circulation of the field in the plane of incidence leads to the spin which is transverse.  In what follows, we briefly discuss all the above cases and identify the origin of the field circulation in each example. The effect of this spin and spin momentum on a particle in terms of the force and torque is then discussed. The chapter ends with a discussion of how one can exploit the coupled systems and perfect absorption to enhance the tiny effects. 

\section{Orbital and spin parts of the momentum \label{sec1-1}}
It is well known from the seminal works of J. H. Poynting that light carries momentum and angular momentum \cite{allenbook,jacksonbook}. These two quantities of light are also preserved in the quantum mechanical picture of photons. An elliptically polarized plane electromagnetic wave in free space propagating along the $z$-direction of Cartesian coordinate system ($xyz$) can be represented as:
\begin{equation}
\mathbf{E}\propto\dfrac{\mathbf{x}+m\mathbf{y}}{\sqrt{1+|m|^2}} exp(ikz),
\label{eq1-1}
\end{equation}  
where $\mathbf{x}$ and $\mathbf{y}$ are unit vectors along the $x$ and $y$ axes, respectively, $m$ defines the polarization state with helicity $\sigma=\dfrac{2Im(m)}{1+|m|^2}$ with  $k$ beging the wave number in free space. The momentum and spin angular momentum (SAM) density of the electromagnetic plane-wave specified by Eq.(\ref{eq1-1}) are given as \cite{bornwolf,jacksonbook}:
\begin{equation}
\mathbf{p}\propto \mathbf{k};\;\;\;\; \mathbf{s}\propto \dfrac{\sigma \mathbf{k}}{k}
\label{eq1-2}
\end{equation}
where $\mathbf{k}=k\mathbf{z}$ is the wave-vector. Eq.(\ref{eq1-2}) agrees with the quantum mechanical picture of photons as the momentum $\mathbf{p}$ is determined by the wave-vector $\mathbf{k}$ independent of polarization and the SAM density is dependent on helicity and it is also collinear with the wave-vector. Note that the momentum and SAM of the field can be realized experimentally by placing a small absorbing particle in the field and observing the force and torque acting on it. The radiation force and torque on such a probe particle are thus proportional to the momentum $(\mathbf{F}\propto\mathbf{p})$ and torque $(\mathbf{T}\propto \mathbf{s})$, respectively \cite{ashkin1983,chaumet2000}.
This picture of momentum and SAM is valid and well-understood for plane-waves (or paraxial beams neglecting the longitudinal component). However, for inhomogeneous optical fields (e.g. evanescent fields), we also get additional transverse components of both momentum and spin densities as discussed below.
\par
The momentum density ($\mathbf{p}$) in a monochromatic optical field can be decomposed into orbital and spin parts such that $\mathbf{p}=\mathbf{p}_s+\mathbf{p}_o$ \cite{berry2009,berry2015,bliokh2014,bliokhphyrep}. The orbital $\mathbf{p}_o$ (canonical) and spin-momentum densities $\mathbf{p}_s$ can be expressed as\cite{aiello2015,bliokh2014,berry2009}:
\begin{eqnarray}
\mathbf{p}_o&=&\dfrac{Im[\epsilon_0 \mathbf{E^*}.(\nabla)\mathbf{E}+\mu_0\mathbf{H^*}.(\nabla)\mathbf{H}]}{4\omega}, \label{eq1-3}\\
\mathbf{p}_s&=&\dfrac{1}{2}\nabla\times\mathbf{s}.
\label{eq1-4}
\end{eqnarray}
In Eq. (\ref{eq1-3}) $\mathbf{s}$ is the SAM density given by \cite{aiello2015,bliokh2014}:
\begin{equation}
\mathbf{s}=\dfrac{1}{4\omega} Im(\epsilon_0 \mathbf{E^*} \times \mathbf{E}+ \mu_0 \mathbf{H^*}\times\mathbf{H})
\label{eq1-5}
\end{equation}
and the operator $\nabla$ is defined by $\mathbf{A}\cdot(\nabla)\mathbf{B} \equiv A_x\nabla B_x+A_y\nabla B_y+A_z\nabla B_z $. Note that in dispersive medium with frequency dependent  $\epsilon(\omega)$ and $\mu(\omega)$, one needs to replace $\epsilon_0$ and $\mu_0$ by $\tilde{\epsilon}=\epsilon+\omega\dfrac{d\epsilon}{d\omega}$ and $\tilde{\mu}=\mu+\omega\dfrac{d\mu}{d\omega}$\cite{bliokh2017prl,bekshaev2018}.
The canonical momentum density ($\mathbf{p}_0$)  is associated with the local wave-vector in a generic optical field and responsible for exerting radiation pressure force\cite{berry2009,bliokh2014}. The spin-momentum density ($\mathbf{p}_S$) has usually been considered to be a virtual quantity in the sense that it does not exert any force on a dipolar particle.
\par
In what follows we show the presence of extraordinary spin and momentum components in a structured optical field. Specifically, we cite the examples of a propagating Gaussian beam and an evanescent field as in total internal reflection.
\subsection*{Propagating Gaussian beam}
In paraxial approximation, the longitudinal and transverse components of a linearly polarized ($x$-polarized) collimated Gaussian beam at the focal plane can be expressed as \cite{aiello2015}:
\begin{eqnarray}
E_x(x,y)&\propto& \dfrac{1}{z_0} exp\big(-\dfrac{k}{2} \dfrac{x^2+y^2}{z_0}\big), \label{eq1-6}\\
E_z(x,y)&\propto& \dfrac{-ix}{z_0^2} exp\big(-\dfrac{k}{2} \dfrac{x^2+y^2}{z_0}\big). \label{eq1-7}
\end{eqnarray}
The following interesting features can be discerned from Eqs. (\ref{eq1-6}) and (\ref{eq1-7}):
\begin{enumerate}
	\item  the longitudinal component $E_z$ is shifted in phase w.r.t. the transverse component $E_x$ by an amount of $\pi/2$ leading to a transverse spinning of the field in region where the two components co-exists.
	\item The symmetric Gaussian profile in $E_x$ leads to asymmetric profile in $E_z$.
	\item The strength of longitudinal component $E_z$ depends on  beam waist $w_0$ parameter ($z_0=\dfrac{kw_0^2}{2}$), thus tightly focused Gaussian beams lead to higher transverse SAM density.
\end{enumerate}
\subsection*{Evanescent field}
\begin{figure}[ht]
	\centering
	\includegraphics[width=0.65\linewidth]{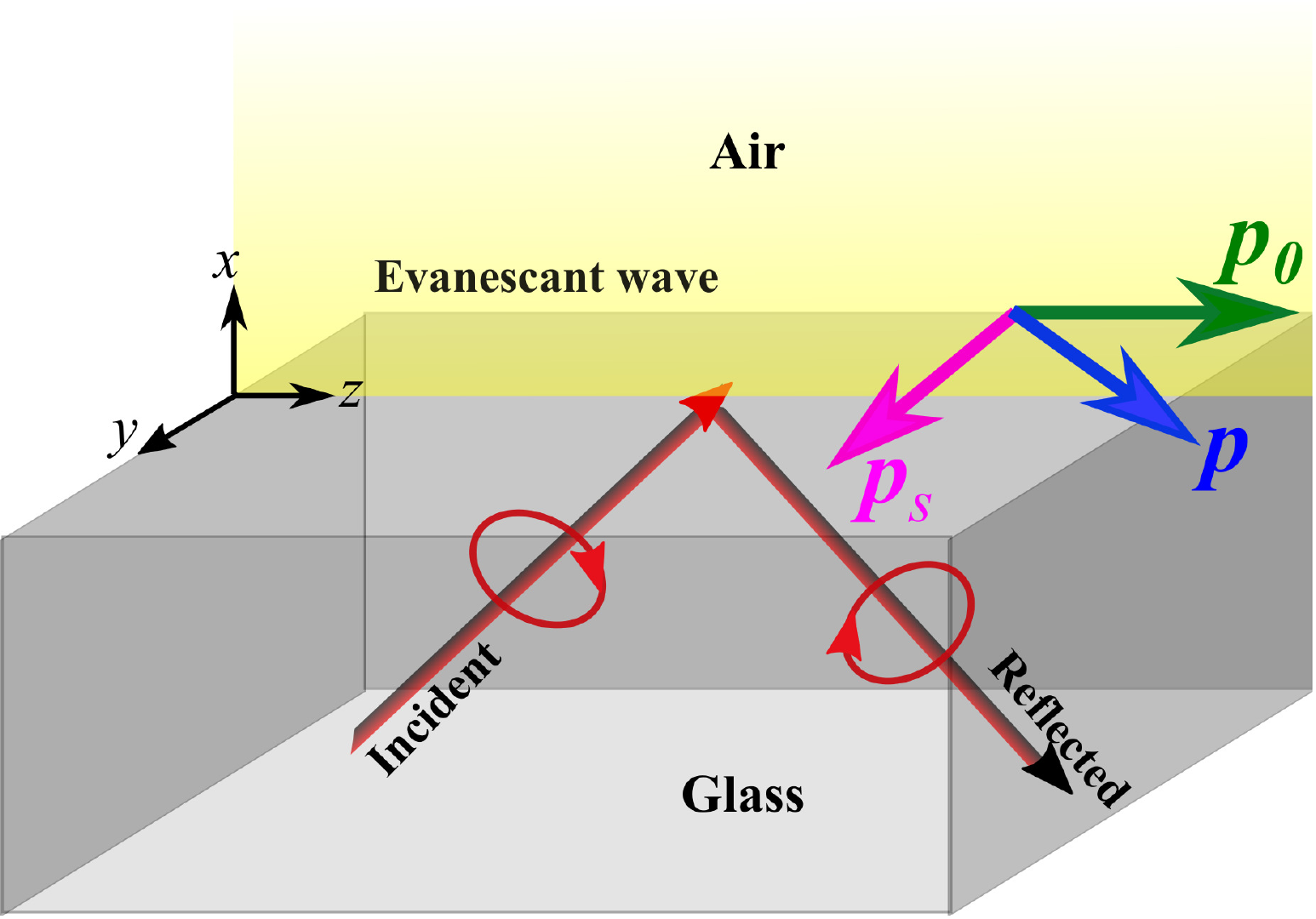}
	\caption{Schematic of the evanescent field generation by total internal reflection of a circularly polarized plane wave incident on a air-glass interface above the critical angle from the $x<0$ half-space. The components of momentum densities of the evanescent field in $x>0$ half space are shown. Here $\mathbf{p}_0$ and $\mathbf{p}_s$ represent respectively the orbital and spin parts of the total momentum density $\mathbf{p}$.    } 
	\label{fig1-1}
\end{figure}
Consider an evanescent field resulting from total internal reflection at an interface between two dielectric media both having real dielectic constants (lossless). The schematic for a glass-air interface for incidence with a polarized (characterized by $m$) light is shown in Fig. \ref{fig1-1}. Let the wave propagate along the $z$-direction and decay in the $x>0$ half space. Such an evanescent wave can be represented mathematically by \cite{bliokh2014,bliokhphyrep}:
\begin{equation}
\mathbf{E} \propto \dfrac{1}{\sqrt{(1+|m|^2)}} \big(\mathbf{x}+m\dfrac{k}{k_z}\mathbf{y}-i\dfrac{\kappa}{k_z}\mathbf{z} \big)  exp(ik_z z-\kappa x)
\label{eq1-8}
\end{equation}
with the complex propagation constant  $\mathbf{k}=k_z\mathbf{z}+i\kappa \mathbf{x}$ such that $k_z^2=\kappa^2+k^2$. One can calculate the canonical momentum, spin momentum and SAM densities for the evanescence field of Eq. (\ref{eq1-8}) and their expressions are found to be \cite{bliokh2014,bliokhphyrep}:
\begin{eqnarray}
\mathbf{p}_o&\propto& Re(\mathbf{k}), \label{eq1-9}\\
\mathbf{p}_s &\propto& -\dfrac{(Im \mathbf{k})^2}{(Re\mathbf{k})^2} Re\mathbf{k}+\sigma k \dfrac{Re\mathbf{k}\times Im \mathbf{k}}{(Re\mathbf{k})^2}, \label{eq1-10}\\
\mathbf{s}&\propto& \Big(\sigma \dfrac{k}{k_z} \mathbf{z}+\dfrac{\kappa}{k_z}\mathbf{y}\Big). \label{eq1-11}
\end{eqnarray}
It is thus clear from Eq. (\ref{eq1-10}) that there is a transverse momentum component which is proportional to the input wave helicity $\sigma$, while Eq. (\ref{eq1-11}) leads to transverse SAM density $s_y$ which is independent of the input wave helicity. These two components of momentum and spin are in sharp contrast to what is known for propagating fields or photons. The presence of the imaginary longitudinal component of  \textbf{E} in Eq.(\ref{eq1-8}) leads to a rotation of the field in the $xz$-plane of the evanescent field. This rotation generates the transverse SAM density and transverse momentum components. Note that the transverse SAM is independent of input polarization, hence even present for a linearly polarized evanescent beam for which $m=\sigma=0$. The mechanical action of this transverse SAM is a torque on a probe particle in the transverse direction ($y$) which may lead to a rotation of the probe particle in the $xz$-plane. From Eq. (\ref{eq1-9}), we see that the canonical part of the momentum density is proportional to the real part of the complex wave-vector of the evanescent field. Eq. (\ref{eq1-10}) indicates that the spin-momentum density has a longitudinal component which is opposite to $\mathbf{p}_0$ and a helicity-dependent unusual transverse component $(\mathbf{p}_s)_y$.
\par
The force experienced by a dipolar Rayleigh particle with equal electric and magnetic polarizability $\alpha_e=\alpha_m=\alpha$ is given by the sum of gradient force and the radiation pressure force and is independent of spin-momentum contribution. This optical force on the dipolar probe particle is given by \cite{ashkin1983,chaumet2000,berry2009,ebbesen2003}:
\begin{equation}
\mathbf{F}\propto \dfrac{1}{2\omega} Re(\alpha)\nabla w + Im (\alpha) \mathbf{p}_0
\label{eq1-12}
\end{equation}
where $w$ is the energy density of the field. The first term in Eq. (\ref{eq1-12}) is the gradient force, while the second term is the canonical force on the particle. Considering higher order interaction in the light matter interactions and calculating the higher-order corrections to the dipole optical force in a non-absorbing particle, the correction term in the optical force is obtained to be $\delta\mathbf{F}\propto -Re(\alpha_e\alpha_m)\mathbf{p}$ \cite{bliokh2014}.  The transverse component of which becomes proportional to $\delta\mathbf{F}y \propto (\mathbf{p}_s)_y \propto \sigma y$\cite {bliokh2014}. This relation thus gives an opportunity to detect the Belinfante spin-momentum ($\mathbf{p}_s$) and the novel helicity-dependent transverse force. This novel transverse force has indeed been measured experimentally using a probe immersed in an evanescent field \cite{natphys_2016}. It may be noted that the transverse force is two-orders of magnitude weaker compared to the longitudinal force rendering the experimental detection to be really challenging.
\par 
From the Eq.\ref{eq1-10}, one can identify an inherent property of the evanescent wave, namely the spin-momentum locking which implies that the direction of propagation of momentum gets locked with the polarization of the input wave. The origin of such spin-momentum locking was shown to be related with the dispersion and causality relations associated with the evanescent waves \cite{zubinoptica,zubinapl}. This property of the evanescent wave was identified to be solely related to the decaying nature of the field and independent of the nanophotonic structure \cite{zubinoptica}. The spin-momentum locking phenomenon in surface optical modes has been identified as the quantum spin Hall effect of light and analogies have been drawn with the topological insulators for electrons \cite{bliokh2015}.
\par
It is pertinent to comment on the clean phase separation by $\pi/2$ between the $x$- and the $z$- components of the field given by Eq. (\ref{eq1-8}). This is a consequence of the nonlossy nature of both the dielectric media. In case of lossy media, especially for metals one can not ignore the imaginary part of the dielectric function and the simple relation does not hold. In particular, for surface plasmons on a metal dielectric interface \cite{raetherbook, sdg1998, sdgbook}, it is essential to retain the finite imaginary part of the dielectric function and the problem needs to be handled numerically especially with the experimental data of Johnson and Christie \cite{johnson1972} for noble metals. Note that the neglect of the imaginary part of the dielectric function amounts to the neglect of the distinguishability of the long and short range modes of a thin metal film. In fact the attributes long and short loose their meaning if one assumes the dielectric function to be real and negative as in the earlier work on transverse spin \cite{bliokh2012}. Nevertheless, neglect of the imaginary part still brings out the exotic physics of transvere spin and spin momentum. See Chapter \ref{chapter2} for a more detailed discussions of these issues. 
\section{Role of dispersion management and avoided crossings on tranverse spin and momentum \label{sec1-2}}
The generic features of two mode coupling for both closed (lossless) and open (dissipative) systems were studied in detail by Wiersig \cite{wiersig2006}. It was shown that these features can be captured by looking at the eigenvalues and eigenvectors of a model $2\times2$ Hamiltonian given by 
\begin{equation}\label{eq1-13}
H=\left(\begin{array}{cc}
E_1	& V \\ 
W	& E_2
\end{array} \right).
\end{equation} 
For a lossless system the matrix is Hermitian with real $E_j (j=1,2)$ and $W=V^*$ and the eigenvalues are given by
\begin{equation}\label{eq1-14}
E_{\pm}=\frac{E_1+E_2}{2}\pm\sqrt{\left(\frac{\Delta E}{2}\right)^2+|V|^2}, ~~~ \Delta E=E_1-E_2.
\end{equation} 
for a finite interaction $V\ne 0$. The dependence of $E_{\pm}$ on $\Delta E$ clearly reveals the normal mode splitting and avoided crossing. It was also shown by Wiersig \cite{wiersig2006} that the same Hamiltonian Eq.(\ref{eq1-13}) describes the behavior of open dissipative systems when the energies $E_j (j=1,2)$ are allowed to be complex. In fact the imaginary part of the energies define the lifetimes of the coupled system. For appropriate system parameters there can be a splitting of the lifetimes leading to subnatural linewidth. 
\par The above mentioned generic features have been observed in different areas of physics ranging from cavity QED to coupled plasmonic structures. For example, light-matter interaction in high finesse cavity was shown to lead to the vacuum-field Rabi splittings  \cite{mondragon1983,agarwal1984,raizen1989,yamamoto1993,vahala2004}. As in the generic case the coupling occurs when the dispersion branches of the uncoupled systems cross. 
We cite few other examples of strong coupling in varied areas highlighting their application potentials. These include 
fast and slow light \cite{shimizu2002,manga2004}, optical sensing \cite{hanOL2010}, counting and sizing of nanoparticles \cite{sahin2010}, planar or spherical plasmonic or guided wave structures, metamaterial cavities, in photonic crystal fibers \cite{sdg1985,sdg1987,sdg1995,surfaceplasmonPR,sdg2009,rohde2007,shourya2010,hanOL2010,jansen2011}, etc. It was believed that the the split modes can be resolved only with high-finesse optical modes. It is now understood that such effects can be observed in systems with substantial losses (eg. plasmonic nanocavities, and leaky cavities)  \cite{reithmaier2004,benz2013,lien2016}. 
\par
Our interest in the strongly coupled systems is dictated by the fact that there is a significant change in the dispersive and absorptive properties near the anti crossing as a consequence of mode mixing and exchange \cite{sdg1995,manga2005, pavan2013}. The obvious consequence is a highly structured field which is essential for observing  the Belinfante's transverse spin \cite{belinfante1940} in optical systems \cite{aiello2015,bliokhphyrep,measuring2015,natphys_2016}. In Chapters \ref{chapter2} and \ref{chapter3}, we discuss the results on the effects of mode coupling on this elusive transverse spin in planar and spherical micro and nano structures, respectively.
\section{Role of coherent perfect absorption in field enhancement \label{sec1-3}}
\begin{figure}[ht]
\begin{center}
\includegraphics[scale=0.3]{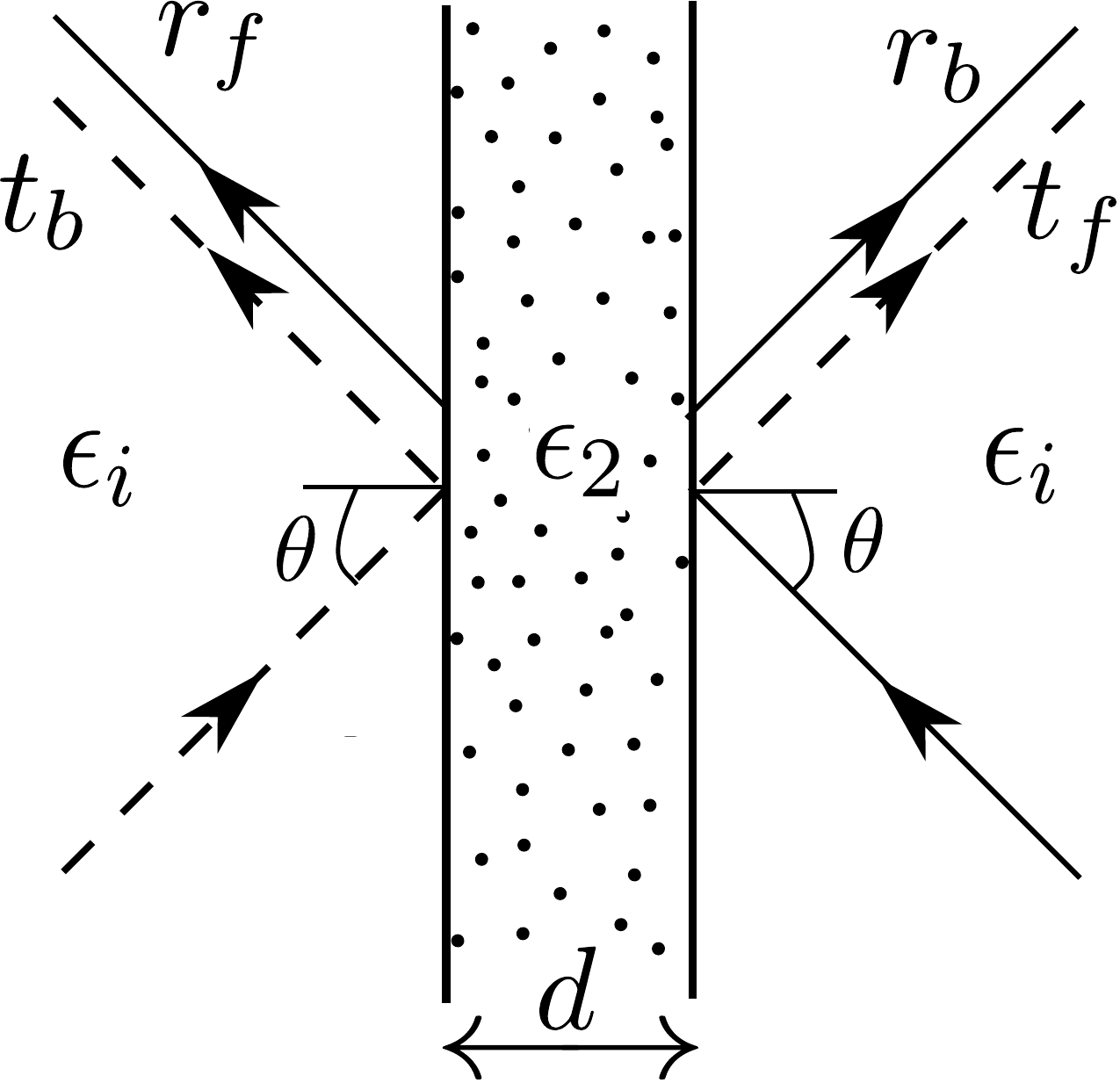}
\caption{Schematics of oblique incidence CPA under plane wave illumination from both sides. The subscript $f$ ($b$) refer to quantities for incidence from left (right). Republished with permission of Taylor and Francis Group LLC Books, from Wave Optics: Basic Concepts and Contemporary Trends,  Subhasish Dutta Gupta, Nirmalya Ghosh, and Ayan Banerjee, CRC Press, 2015; permission conveyed through Copyright Clearance Center, Inc. }
		\label{fig1-2}
	\end{center}
\end{figure}
Coherent perfect absorption (CPA) and its predecessor critical coupling (CC) have drawn considerable attention in optics research in recent years \cite{wan2011,dutta-gupta12,dutta-gupta2012,yarivcc,cai2000,sdg2007}. Both rely on coherent illumination and perfect destructive interference with the same goal of transferring all the incident light energy in the system. While CPA uses illumination from both sides, CC is characterized by illumination from one side. Suppression of scattering in CPA is achieved by destructive cancellation on both sides of the system, while in CC transmission is prohibited by a Bragg reflector or a metal layer with destructive cancellation of the reflected light. A detailed description of both CPA and CC can be found in reference \cite{sdgbook}. Recently the theory has been extended to cover nonlinear systems as well \cite{nireekshan2013,nireekshan2014,li2018}. It has also been exploited to explore the possibility of perfect visibility in two-photon quantum interference \cite{sdg2014}, where selective excitation of one of the coupled modes was used.
\par
The general oblique incidence scenario for CPA is shown in Fig.~\ref{fig1-2}. CPA is achieved when amplitude reflection coefficient matches the amplitude transmission coefficient in magnitude with a relative phase difference of $\pi$ (i.e. when $|r_f|=|t_b|$ and the relative phase $\delta \phi_{r-t}=\pi $, with analogous relations on the other side).  The first CPA mediated manipulation of absorption was experimentally reported by Wan  \textit{et al.} \cite{wan2011}. In recent years it has been exploited for many other interesting effects \cite{pu2012,roger2015,wu2016,zhu2017,zhu2016}. 
\par
In the context of transverse spin the CPA effect with one of the preselected modes of a strongly coupled system can play a major role. It stems from the fact that there is a considerable enhancement of the local fields since all the incident energy is transferred to the preselected mode. Coupled with dispersion management near the avoided crossing, this can lead to enhanced structured fields resulting in resonantly enhanced transverse spin. The discussion on CPA and the associated field enhancement  in a gap plasmon guide is presented in Chapter \ref{chapter2}. Subsequent discussions show the possibility of scaling up the transverse spin.
\chapter{Transverse spin and momentum in planar structures \label{chapter2}}
The simplest possible structures that can support transverse spin and transverse spin-momentum are perhaps the planar structures supporting guided and surface modes. It is quite natural that the first reports of transverse spin was based on optical interfaces supporting evanescent waves under total internal reflection or surface plasmons (see Section \ref{sec1-1}). It was mentioned earlier that the transverse spin associated with evanescent waves is helicity independent while the transverse Belinfante spin momentum has helicity dependence \cite{bliokh2014}. In this Section we focus on planar structures supporting coupled surface plasmons. As clearly stated in Section \ref{sec1-2}, coupling of the surface plasmons offers an additional handle on controlling the transverse SAM by means of engineered dispersion mediated by the strong coupling. This in particular cases can lead to enhanced transverse SAM densities which is essential for easier experimental detection of this elusive extraordinary effect. One of the first studies on transverse spin with surface plasmons  \cite{bliokh2012} ignored the essentially complex character of the dielectric function of metal treating it to be real and negative to bring out the novel physical effects. Full numerical results including the imaginary part of metal dielectric function was treated later by others \cite{mukherjee2017,mukherjee2016} for Sarid configuration and the gap plasmon guide, respectively. However, these studies ignored the dispersion effects \cite{bliokh2017prl,saha2018} defined by the frequency derivatives of the dielectric functions (see discussion after Eq. (\ref{eq1-5})) in metal while calculating the transverse SAM densities (see Section \ref{sec3-1} for an assessment of the contributions from such terms to SAM density).
\section{Transverse spin in Sarid Geometry \label{sec2-1}}
\begin{figure}[t]
	\centering
	\includegraphics[width=0.4\linewidth]{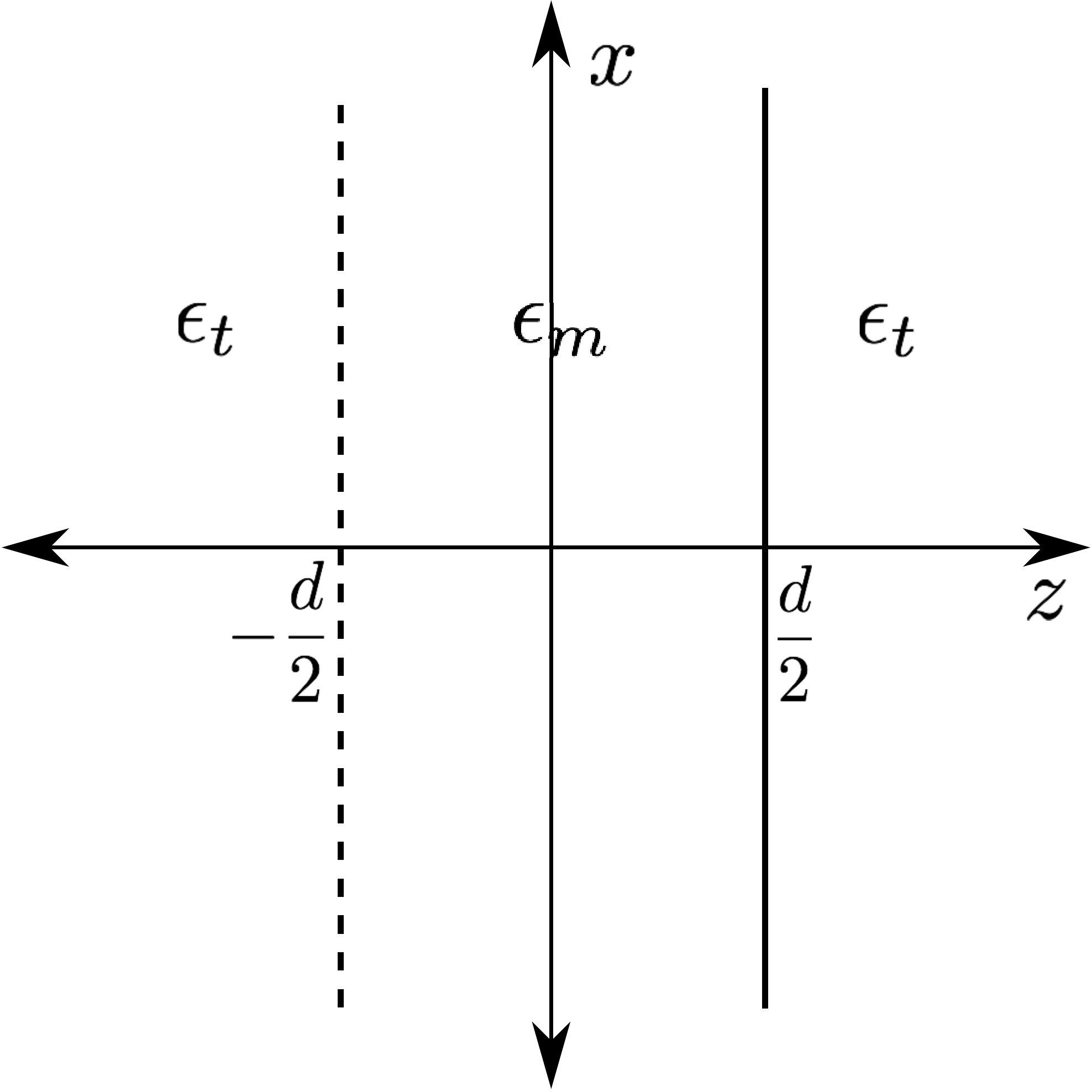}
	\caption{Schematic view metal film with thickness $d$ and dielectric function $\epsilon_m$ in a dielectric with dielectric constant $\epsilon_t$ . Symmetry is exploited reducing the calculations for only the right half of the structure. Reprinted by permission from Springer Nature: Pramana - J. Phys., Transverse spin with coupled plasmons, S. Mukherjee, A. V. Gopal, and S. D. Gupta, (2017).}
	\label{fig2-1}
\end{figure}
\begin{figure}[h]
	\centering
	\includegraphics[width=0.8\linewidth]{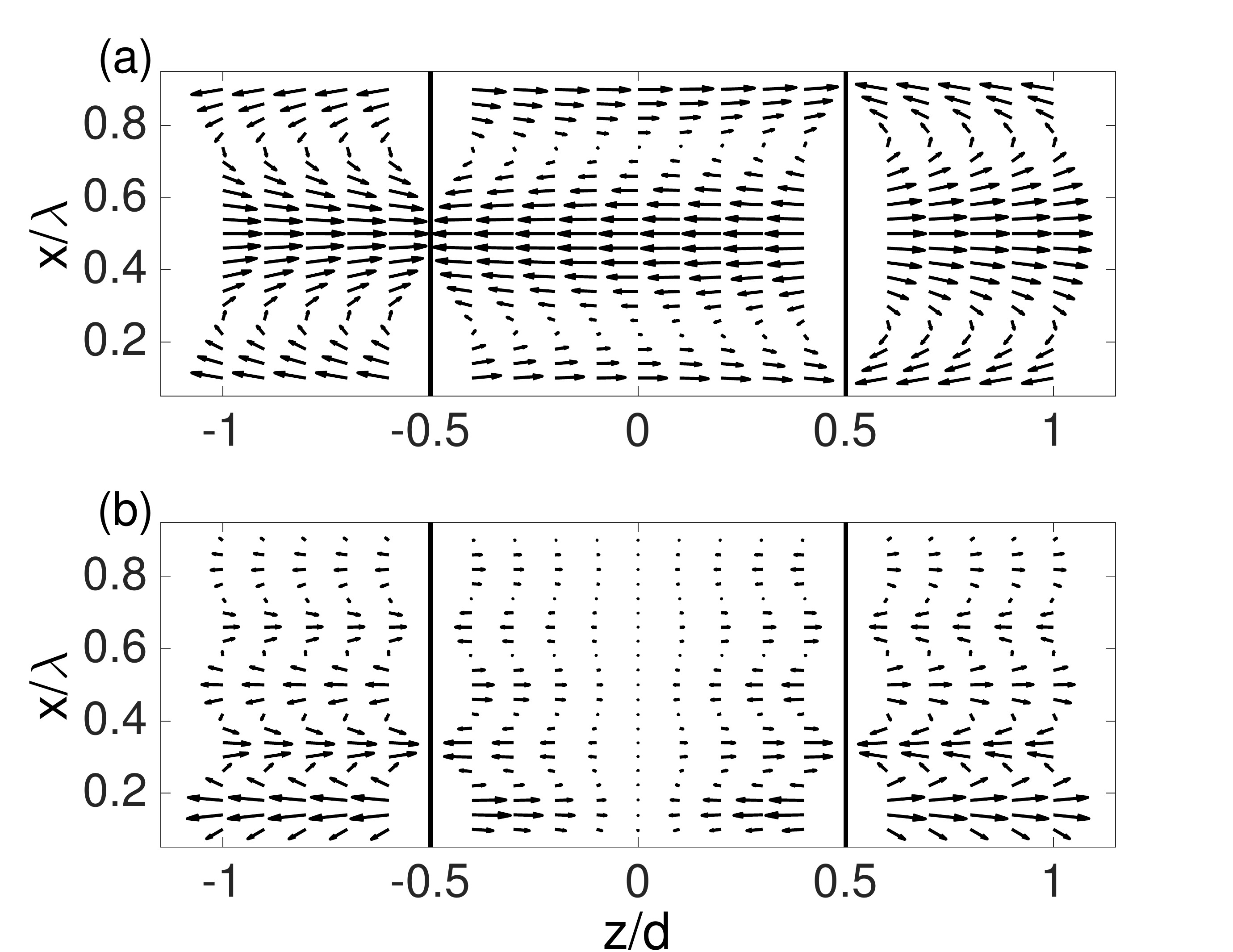}
	\caption{Electric Field in the $x-z$ plane for $\lambda = 0.633\ \mu$m and $d = 0.006\ \mu$m for (a) symmetric and (b) antisymmetric magnetic field profiles, respectively. Reprinted by permission from Springer Nature: Pramana - J. Phys., Transverse spin with coupled plasmons, S. Mukherjee, A. V.  Gopal, and S. D. Gupta, (2017).}
	\label{fig2-2}
\end{figure}

A typical Sarid geometry deals with the coupled modes of a very thin metal film embedded in a dielectric \cite{sarid1982,sdg1987,sdg1998,sdgbook} (see Fig. \ref{fig2-1}). The coupling leads to the long- and short- range modes. Note that the notion of long- and short- range modes are meaningful only in presence of finite imaginary part of the metal dielectric function. The long- range modes are associated with larger quality factor and significant field enhancements, which has been exploited over the years for sensing, surface enhanced nonlinear optical effects and spectroscopy \cite{sdg1998}. The eigenmodes and the dispersion features for the structure shown in Fig. \ref{fig2-1} have been studied extensively \cite{raetherbook,sdg1998,sdgbook}. Because of the inherent symmetry of the structure there can be both the symmetric and the antisymmetric modes (with respect to magnetic fields). For example, for the symmetric case, the dispersion relation and the field amplitude in the dielectric  ($z \ge d/2$) are given by \cite{mukherjee2017}
\begin{eqnarray}
\epsilon_{m}k_{tz} + \epsilon_{t}k_{mz} \tanh\left(\frac{\bar{k}_{mz}d}{2}\right) &= &0\ ,\label{eq2-1}\\
A_{t}^{s} &= &2 A_0 \cosh\left(\frac{\bar{k}_{mz}d}{2}\right)\ .\label{eq2-2}
\end{eqnarray}
Analogous expressions for the antisymmetric modes are given by
\begin{eqnarray}
\epsilon_{m}k_{tz} + \epsilon_{t}k_{mz} \coth\left(\frac{\bar{k}_{mz}d}{2}\right) &=& 0\ ,\label{eq2-3}\\
A_{t}^{a} &=& -2 A_0 \sinh\left(\frac{\bar{k}_{mz}d}{2}\right)\ .\label{eq2-4}
\end{eqnarray}
In Eqs. (\ref{eq2-1})--(\ref{eq2-4}), $k_{mz} = i\sqrt{k_{x}^{2} - k_{0}^{2}\epsilon_{m}}= i\bar{k}_{mz}$, $\epsilon_m$ and $\epsilon_t$ are the dielectric constants of the metal and the dielectric, respectively. Hereafter superscript $s$($a$) denote symmetric (antisymmetric) modes. Using Eqs. (\ref{eq2-1})-(\ref{eq2-4}) one can easily arrive at the field profiles for both the magnetic and electric fields. Furthermore, these can be used in  Eqs.(\ref{eq1-3})-(\ref{eq1-5}) to obtain the transverse spin and transverse spin momentum in a straight forward manner \cite{mukherjee2017}. The circulation of the electric field in the $x-z$ plane for the symmetric and antisymmetric modes is shown in Fig. \ref{fig2-2}. Obviously this circulation is the origin of the transverse spin for the structured light with the coupled modes. Moreover this is more pronounced for the antisymmetric (short- range) modes since the field is more "structured" due to larger dissipation.

\par 
In order to appreciate the circulation of the fields shown in Fig. \ref{fig2-2} , analytical expressions of the fields were evaluated in Reference \cite{mukherjee2017} in the case when $Im(\epsilon_m)=0$. For example, the electric field in the dielectric  were shown to be given by 
\begin{eqnarray}
\mathbf{E}_d^s &=& \frac{2 A_0 \cosh\left(\frac{\bar{k}_{mz}d}{2}\right)}{\omega\epsilon_{t}\epsilon_{0}} \left[i\bar{k}_{tz}\hat{x} - k_x\hat{z}\right]e^{ik_x x}e^{-\bar{k}_{tz}\left(z-\frac{d}{2}\right)}\label{eq2-5},\\
\mathbf{E}_d^a&=& \frac{2 A_0 \sinh\left(\frac{\bar{k}_{mz}d}{2}\right)}{\omega\epsilon_{t}\epsilon_{0}} \left[-i\bar{k}_{tz}\hat{x} + k_x\hat{z}\right]e^{ik_x x}e^{-\bar{k}_{tz}\left(z-\frac{d}{2}\right)}\label{eq2-6},
\end{eqnarray}
which led to the expressions for the transverse spin as follows
\begin{eqnarray}
\mathbf{s}_d^s &=& \frac{-2|A_0|^2 \cosh^2\left(\frac{\bar{k}_{mz}d}{2}\right)}{\omega^{3}\epsilon_{0}\epsilon_{t}^{2}} k_x \bar{k}_{tz}e^{-2\bar{k}_{tz}\left(z-\frac{d}{2}\right)}\hat{y},\label{eq2-7}\\
\mathbf{s}_d^a &=& \frac{-2|A_0|^2\sinh^2\left(\frac{\bar{k}_{mz}d}{2}\right)}{\omega^{3}\epsilon_{0}\epsilon_{t}^{2}} k_x \bar{k}_{tz}e^{-2\bar{k}_{tz}\left(z-\frac{d}{2}\right)}\hat{y}.\label{eq2-8}
\end{eqnarray}
It is noteworthy that for large film thickness $d$, both $cosh$ and $sinh$ approach 1 and Eqs. (\ref{eq2-1}) and (\ref{eq2-4}) yield the dispersion equation for a single interface surface plasmon. Moreover, Eqs. (\ref{eq2-5})--(\ref{eq2-8}) yield the earlier known results on transverse spin density. The phase lag between the $x$ and the $z$ components of the electric field results in the rotation of the field vector in the $x-z$ plane yielding the transverse spin along the $y$ direction.
\section{Transverse spin in gap plasmons \label{sec2-2}}
\begin{figure}[t]
	\centering
	\includegraphics[width=0.6\linewidth]{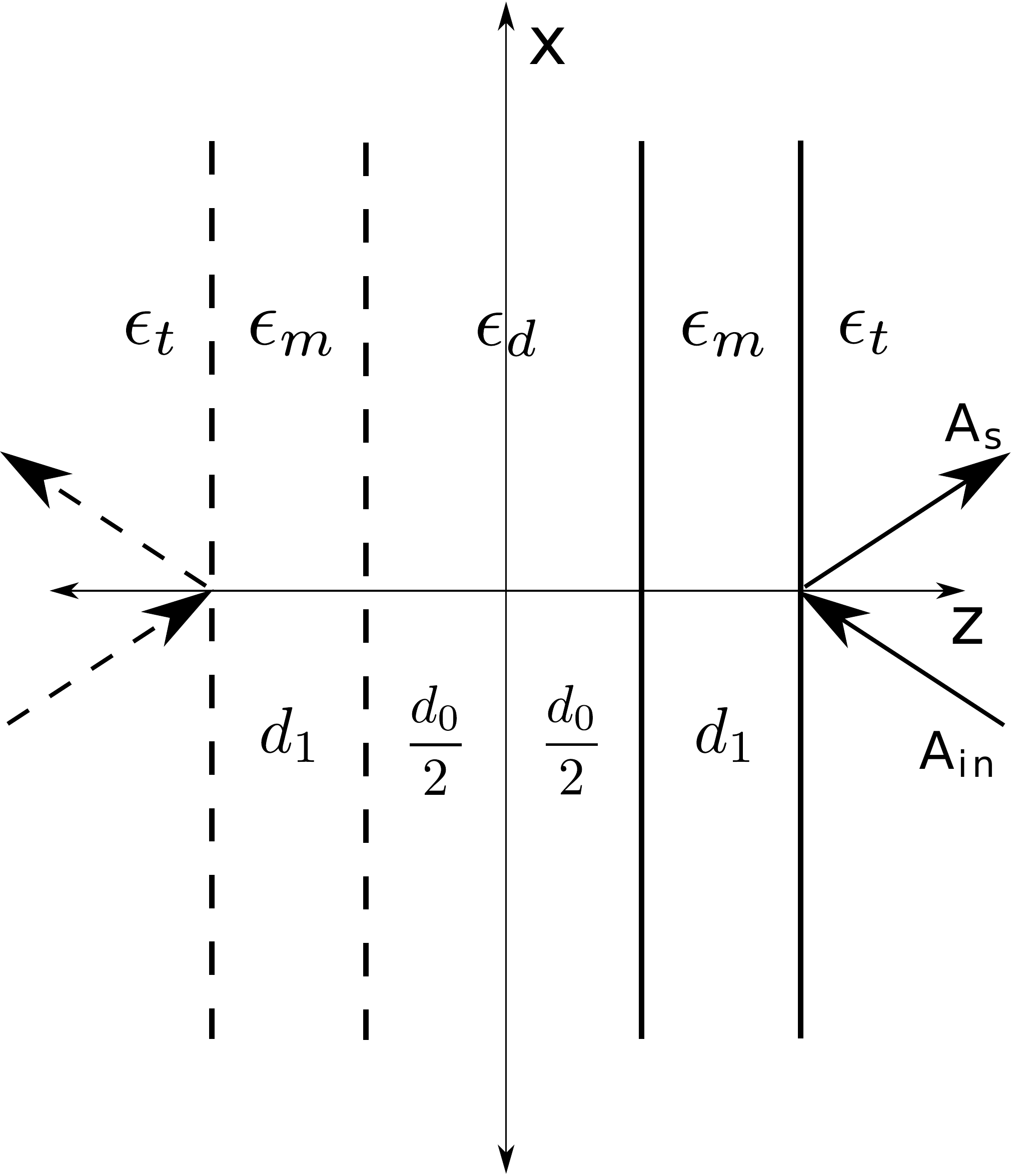}
	\caption{Schematic view of the symmetric gap plasmon guide. We exploit the symmetry and calculate only for the right half of the structure. Adapted from Ref. \cite{mukherjee2016}.}
	\label{fig2-3}
\end{figure}
A typical symmetric gap plasmon guide (two metal films of thickness $d_1$ and  dielectric constant $\epsilon_m$ separated by a dielectric gap of thickness $d_0$ and dielectric constant $\epsilon_d$) is shown in Fig. \ref{fig2-3}. The ambient medium is assumed to have a dielectric permittivity $\epsilon_t$. In order to extract the benefits of CPA let the guide be illuminated from both sides by plane polarized light (say, $TM$) at an angle $\theta$. In contrast to the coupled modes of a thin metal film the modes of a gap plasmon guide offer a better flexibility in the context of dispersion management leading to a variety of interesting effects. These include the nonlinearity induced modes and multistability, possibility of perfect Hong-Ou-Mandel dip and slow light \cite{pande1990,sdg2014,sdg2009}, to name a few. The electric field in the dielectric layer for the symmetric and the antisymmetric modes for the guide shown in Fig. \ref{fig2-3} are given by \cite{mukherjee2016}:
\begin{eqnarray}
\mathbf{\tilde{E}}_d^s = \frac{\tilde{A}_0}{k_0\epsilon_{d}} \left[-2i\bar{k}_{dz} \sinh\left(\bar{k}_{dz}z\right)\hat{x} - 2k_x \cosh\left(\bar{k}_{dz}z\right)\hat{z}\right] e^{ik_x x},\label{eq2-9}\\
\mathbf{\tilde{E}}_d^a = \frac{\tilde{A}_0}{k_0\epsilon_{d}} \left[-2i\bar{k}_{dz} \cosh\left(\bar{k}_{dz}z\right)\hat{x} + 2k_x \sinh\left(\bar{k}_{dz}z\right)\hat{z}\right]e^{ik_x x},\label{eq2-10}
\end{eqnarray}
which clearly shows the already familiar phase lag between the $x$ and the $z$ components of the field. Eqs. (\ref{eq2-9}) and (\ref{eq2-10}) easily lead to the results for transverse spin and the SAM density. 
\begin{figure}[t]
	\centering
	\includegraphics[width=0.6\linewidth]{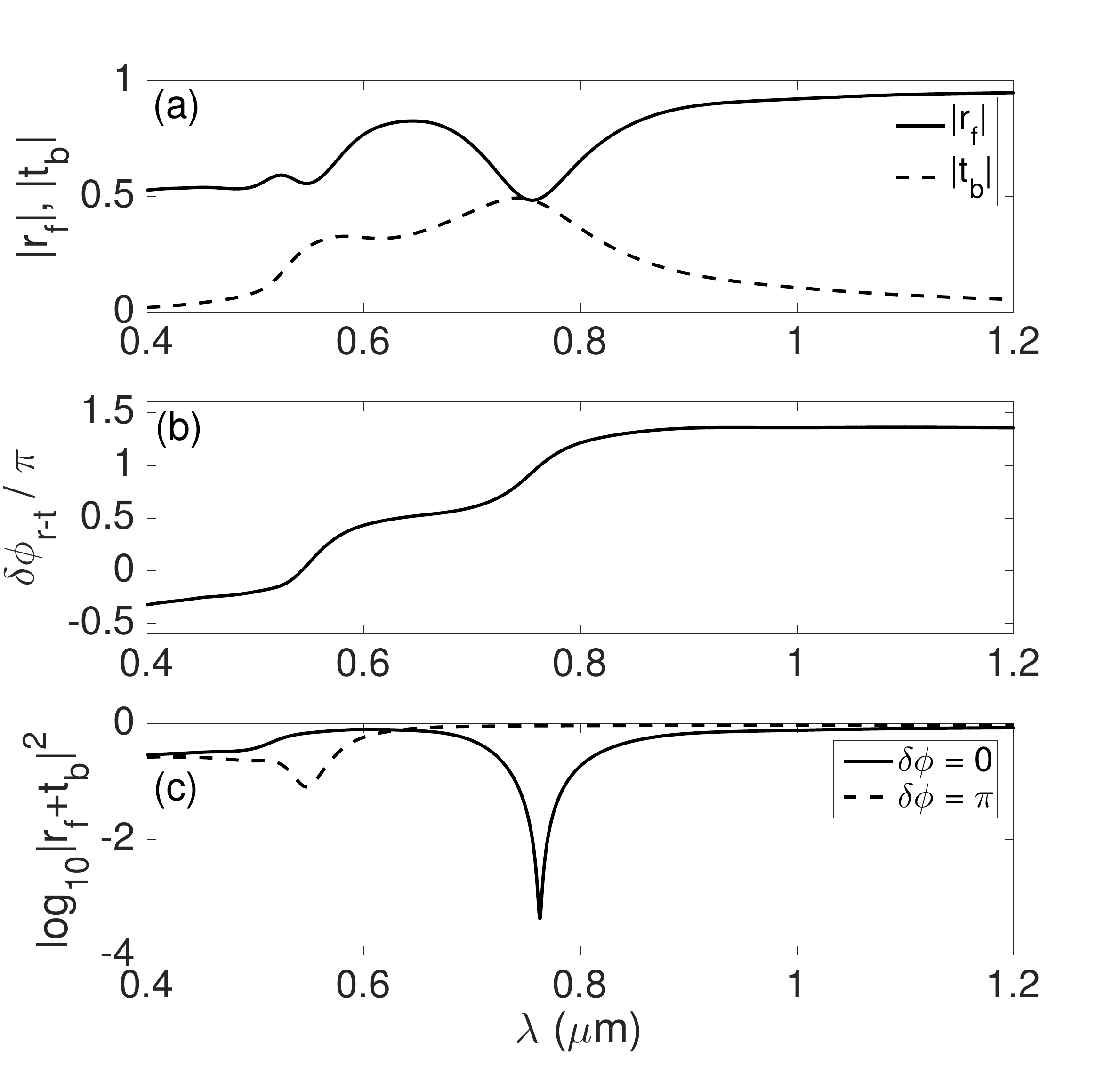}
	\caption{(a) Wavelength dependence of amplitude reflection (transmission) coefficient $|r_f|$ ($|t_b|$), (b) phase difference $\delta \phi_{r-t}$ in units of $\pi$ and (c) the total scattering for the gap plasmon guide with $d_0 = 0.473\ \mu m$, $d_1 = 0.05\ \mu m$ and $\theta = \frac{\pi}{4}$. Other parameters are $\epsilon_d =1$,  $\epsilon_t = 2.28$, $\epsilon_m$ is taken from the work of Johnson and Christy \cite{johnson1972}. Adapted from Ref. \cite{mukherjee2016}.}			
	\label{fig2-4}
\end{figure}
The advantages of the CPA illumination geometry was investigated in detail in reference \cite{mukherjee2016}. We cite the results pertaining to the CPA mediated suppression of total scattering in Fig. \ref{fig2-4}. It is clear that when the (amplitude) reflection  matches the transmission and they are anti-phased, one has the CPA dip. It was shown that there is significant field enhancement at the CPA dip since all the incident field energy is pumped into the coupled mode (see Section \ref{sec1-3} ). The circulation of the electric field vector is shown in Fig. \ref{fig2-5} for both the symmetric and the antisymmetric modes. CPA can facilitate total energy transfer to the preselected long (symmetric) or the short range (antisymmetric) mode.  Fig. \ref{fig2-6} shows the CPA mediated enhanced transverse spin corresponding to these cases.
\begin{figure}[h]
	\centering
	\includegraphics[width=0.6\linewidth]{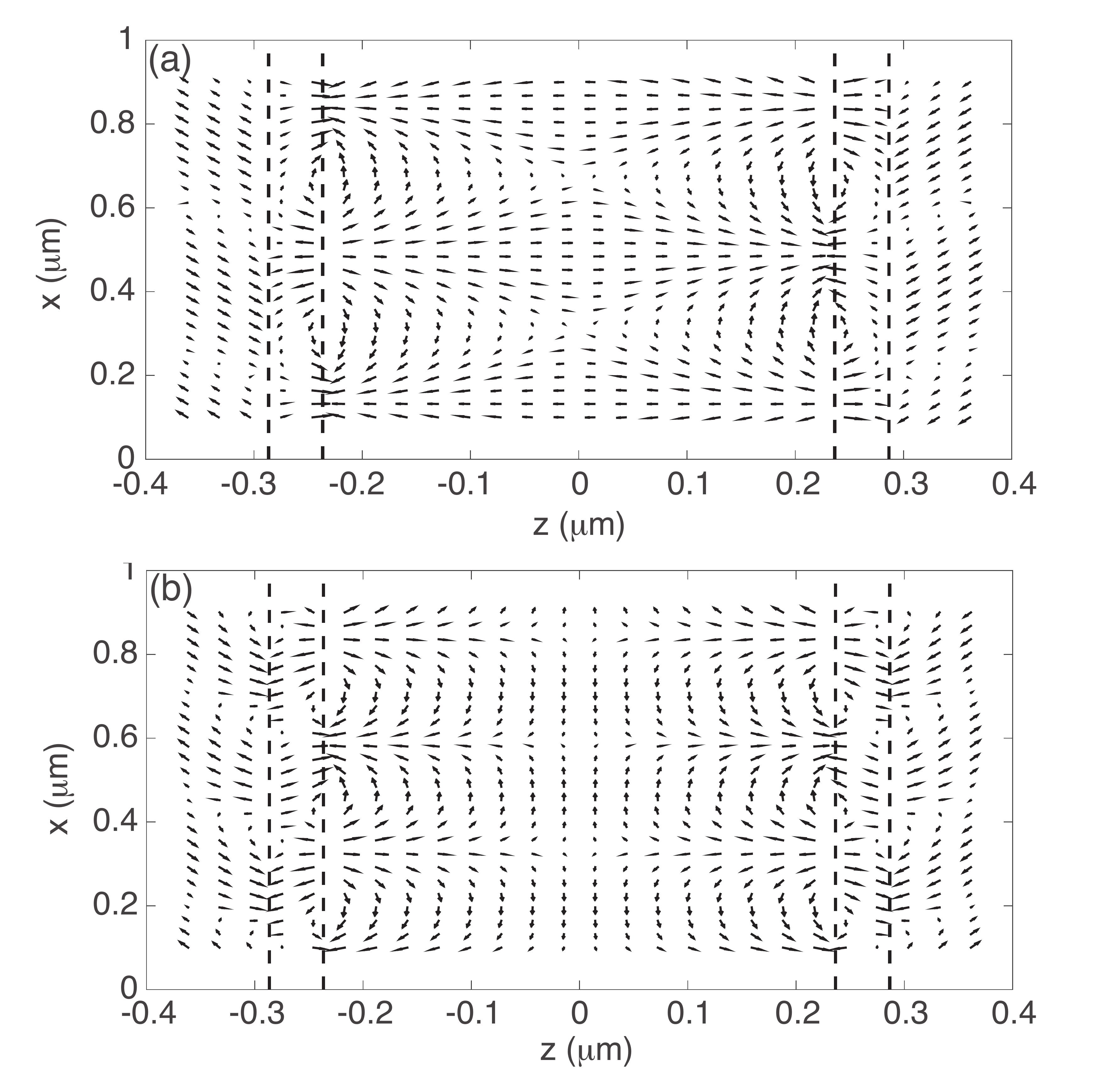} 
	\caption{Electric field quiver plot (not to scale) for (a) symmetric and (b) antisymmetric mode at corresponding excitation wavelengths (i.e., at $\lambda = 0.7628\ \mu$m and at $\lambda = 0.5472~ \mu$m, respectively). Vertical lines represent the interfaces between media. Adapted from Ref. \cite{mukherjee2016}.}
	\label{fig2-5}
\end{figure}
\begin{figure}[t]
	\centering
	\includegraphics[width=1.0\linewidth]{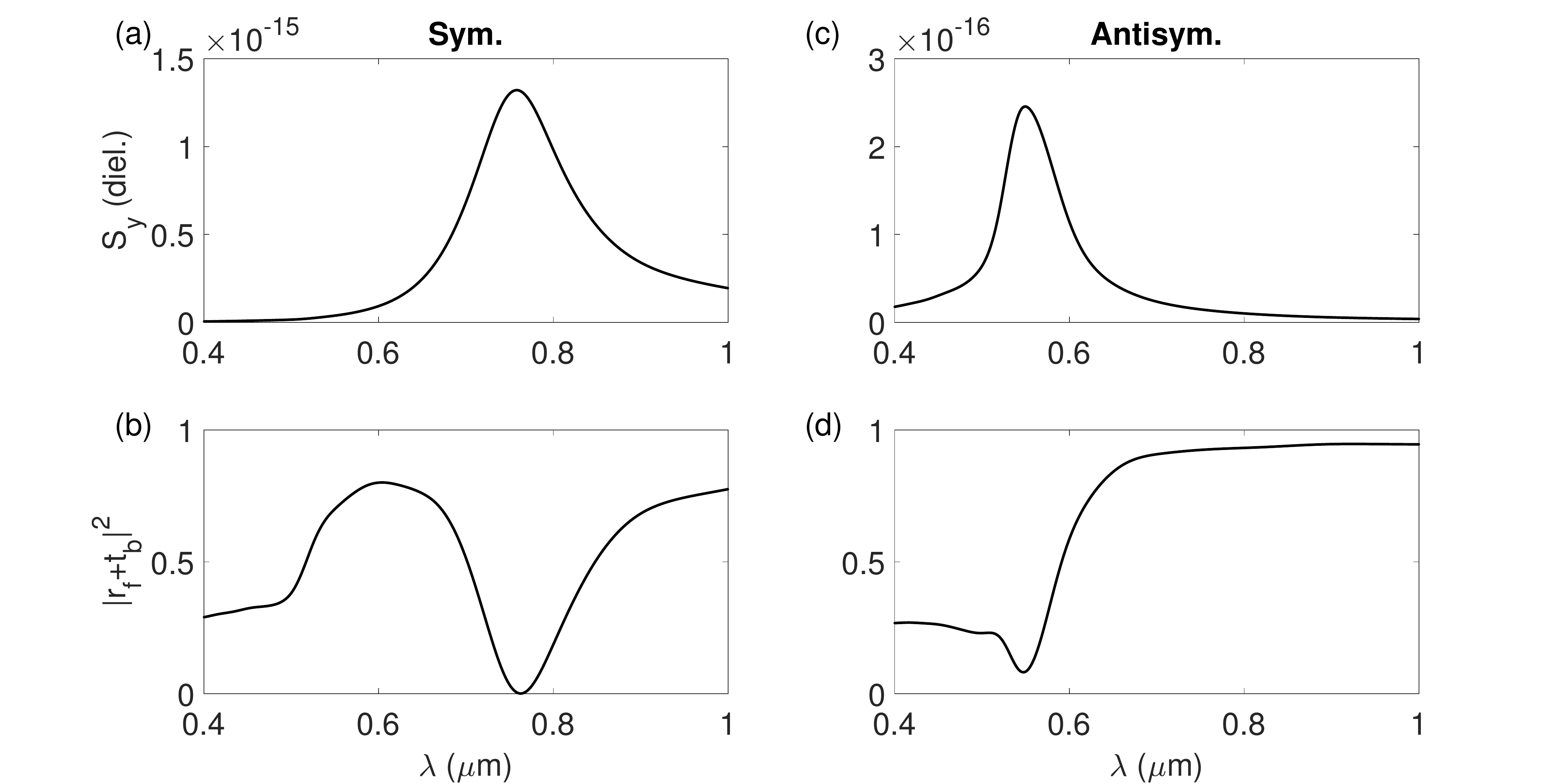}
	\caption{Wavelength dependence of transverse spin at the interface on the dielectric side for the symmetric and the antisymmetric mode. For reference the CPA dip is reproduced from Fig. \ref{fig2-4}. Note the enhanced spin density near the CPA dip. Adapted from Ref. \cite{mukherjee2016}. }
	\label{fig2-6}
\end{figure}

\chapter{Transverse spin and momentum in spherical micro and nano structures \label{chapter3}}
In the preceding Chapters, we have discussed the origin of the transverse SAM and transverse spin momentum in structured optical fields. It appears that finite longitudinal components of the electric and the magnetic fields which are phase shifted with respect to the transverse field components are the origin of both the transverse SAM and the transverse spin momentum of light. It is well known that the general solution of scattering of plane waves from spherical particles also yields such phase-shifted longitudinal component of the scattered electric and the magnetic fields (contributions of which are prominent in the near field and decay rapidly with propagation distance in the far field) \cite{bohren}. The question therefore arises – does scattering of plane waves produce such highly nontrivial structure of the momentum and spin densities? In this Chapter, we address this issue and show that indeed scattering leads to the generation of helicity-independent transverse SAM and polarization-dependent transverse momentum components.
\par
We analyze the influence of the localized plasmon resonances in metal nanosphere and the higher order transverse electric (TE) and transverse magnetic (TM) modes of dielectric microsphere on the near field to far field evolution of the transverse SAM and transverse momentum components of scattered light. We also demonstrate that the interference of the transverse electric and transverse magnetic scattering modes leads to enhancement of both the magnitudes and the spatial extent of the transverse SAM and the transverse momentum components. In this regard, we also discuss the effect of mode coupling in layered spherical scatterers on the transverse spin and momentum densities of the scattered field. 
\section{Transverse SAM and momentum densities in scattering of plane waves from a spherical particle}\label{sec3-1}
\begin{figure}[ht]
	\centering
	\includegraphics[width=0.5\linewidth]{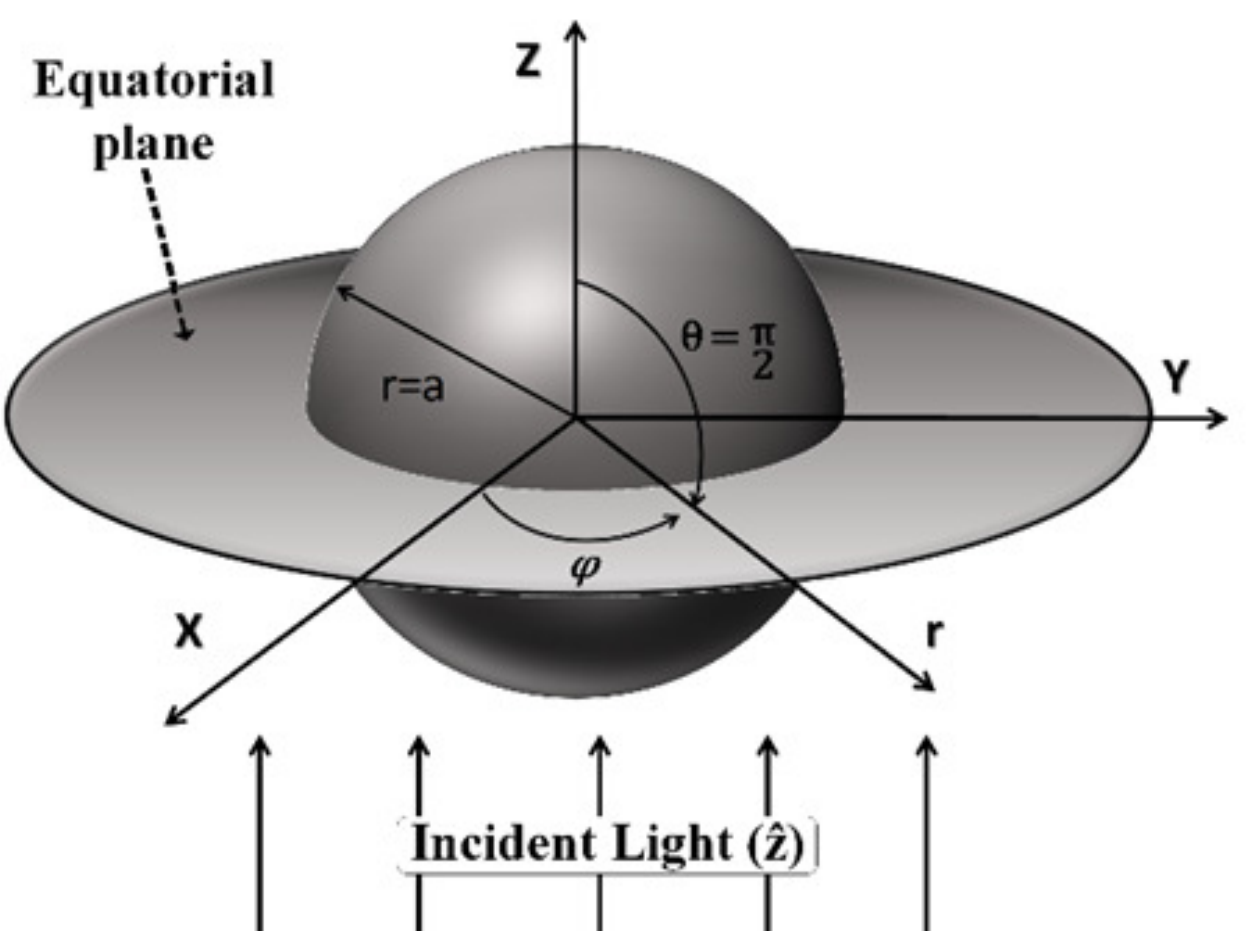}
	\caption{Geometry of problem: An incident plane wave of arbitrary polarization propagating along the $+z$-axis  is incident on a sphere of  radius $a$ centered at the origin of the Cartesian co-ordinate system ($xyz$). The equatorial plane is defined by the polar angle $\theta=\pi/2$ at which the relevant quantities are shown for ease of interpretation. Adapted from Ref. \cite{spiesaha2018}.}
	\label{fig3-1}
\end{figure}
The scattering of plane waves from spherical particles can be described in the framework of standard Mie theory. The expressions for scattered fields for a linearly polarized (say, $x$-polarized) plane wave incident on a homogeneous spherical scatterer of radius $a$ can be written as \cite{bohren}:
\begin{eqnarray}
\mathbf{E}_s&=&\sum_{n=1}^{\infty} E_n (ia_n \mathbf{N}^{(3)}_{e1n}-b_n \mathbf{M}^{(3)}_{o1n}), \label{eq3-1} \\
\mathbf{H}_s&=&\dfrac{k}{\omega\mu}\sum_{n=1}^{\infty} E_n (ib_n \mathbf{N}^{(3)}_{o1n}+a_n \mathbf{M}^{(3)}_{e1n}),
\label{eq3-2}
\end{eqnarray}
where $E_n= i^n E_0 \dfrac{(2n+1)}{n(n+1)}$ with $E_0$ being the amplitude of the incident plane wave, $a_n$ and $b_n$ are the transverse magnetic and transverse electric scattering amplitudes, $\mathbf{M}$ and $\mathbf{N}$ are vector spherical harmonics which satisfy the Helmholtz equations \cite{bohren}. The Mie modes are defined by the poles of the corresponding scattering amplitudes. In general, the scattered electric and magnetic fields have all the three components $(\hat{r},\hat{\theta},\hat{\phi})$ in spherical polar geometry. The radial component is along the direction of propagation of the scattered wave (along $\hat{r}$), the angular component (specified by $\hat{\theta}$) lies in the scattering plane and it is orthogonal to $\hat{r}$, the azimuthal component (specified by $\hat{\phi}$) is orthogonal to the scattering plane. For a resonant TM ($a_n$) mode, the scattered electric field has a longitudinal ($\hat{r}$) component but the magnetic field has no longitudinal component and the reverse is the case for a resonant TE ($b_n$) mode (the vector spherical harmonics $\mathbf{N}$ has radial component but $\mathbf{M}$ does not possess any radial component). Using symmetry of the scattering problem, one can find the scattered fields for input linear $y$-polarized plane wave and for incident right and left circularly polarized (RCP/LCP, $\sigma=\pm 1$) plane waves \cite{spiesaha2018}. These expressions for the scattered fields can then be utilized to obtain the normalized SAM density ($\mathbf{S}$) and the Poynting vector ($\mathbf{P}$) of the scattered wave as \cite{bliokhphyrep,aiello2015,bekshaev2015,bliokh2017njp,bliokh2017prl,bauer2016,bekshaev2018}:
\par
\begin{eqnarray}
\mathbf{S}&=&\dfrac{Im(\tilde{\epsilon} \mathbf{E^*} \times \mathbf{E}+\tilde{\mu} \mathbf{H^*} \times \mathbf{H})}{\omega (\tilde{\epsilon} |\mathbf{E}|^2+\tilde{\mu} |\mathbf{H}|^2)} \label{eq3-3}, \\
\mathbf{P}&=&\dfrac{1}{2} Re(\mathbf{E}\times\mathbf{H}^*) \label{eq3-4},
\end{eqnarray}
where $\tilde{\epsilon}=\epsilon+\omega \frac{d\epsilon}{d\omega}$ and $\tilde{\mu}=\mu+\omega \frac{d\mu}{d\omega}$ are the permittivity and permeability for a dispersive medium, respectively, while for non-dispersive medium  $\tilde{\epsilon}=\epsilon$ and $\tilde{\mu}=\mu$. Note that in contrast to Eq. (\ref{eq1-5}),  Eq. (\ref{eq3-3}) includes a normalization by the energy density.
\par
Clearly, for the three dimensional scattered fields, $\mathbf{S}$ and $\mathbf{P}$ comprise of the radial ($S_r$,$P_r$), angular ($S_\theta$,$P_\theta$) and the azimuthal ($S_\phi$,$P_\phi$) components. Irrespective of the scattering system and its composition, for incident LCP / RCP light, they were shown to satisfy the following relations \cite{saha2016,saha2018}:
\begin{align}
(S_r)_{RCP}&=-(S_r)_{LCP},~~~~~~ &&(P_r)_{RCP}=(P_r)_{LCP}, \nonumber \\
(S_\theta)_{RCP}&=-(S_\theta)_{LCP},~~~~~~ &&(P_\theta)_{RCP}=(P_\theta)_{LCP}, \nonumber  \\
(S_\phi)_{RCP}&=(S_\phi)_{LCP},~~~~~~&& (P_\phi)_{RCP}=-(P_\phi)_{LCP}.
\label{eq3-5}
\end{align}
The radial components of $\mathbf{S}$ and $\mathbf{P}$ represent the conventional longitudinal SAM density and longitudinal momentum, respectively, satisfying the usual dependence on the input helicity of the wave (while longitudinal SAM is opposite for $\sigma=\pm 1$, the longitudinal momentum is independent of $\sigma$). The angular components (perpendicular to the propagation direction but lying in the plane of scattering) also show this usual behavior. In contrast, the azimuthal components exhibit unusual characteristics, while $S_\phi$ is independent of the helicity, $P_\phi$ depends upon the input circular polarization state. Accordingly, $S_\phi$ and $P_\phi$ can be interpreted as the helicity-independent transverse SAM and helicity-dependent transverse (spin) momentum, respectively
\cite{saha2016}. It is pertinent to note that the SAM density (Eq. (\ref{eq3-3})) has both electric and magnetic contributions ($\mathbf{S}=\mathbf{S}^e+\mathbf{S}^m$), which are equivalent for plane waves or for paraxial waves ($\mathbf{S}^e = \mathbf{S}^m$) \cite{berry2009,bliokh2014}. However, in case of the scattered fields, the two contributions can differ significantly. This follows from the fact that the longitudinal component of the scattered field can either be purely electric (for the resonant TM- $a_n$ modes) or purely magnetic (for the resonant TE- $b_n$ modes) in nature. Accordingly, the transverse SAM density ($S_\phi$) would either have electric (TM mode) or magnetic contributions (TE mode). In general, for non-resonant scatterers (having contributions of both the TM and TE modes), $S_\phi$ will have both electric and magnetic contributions, the relative contributions being determined by the strength of the contributing modes and their interference. It may also be worth mentioning that the angular ($P_\theta$) and the azimuthal ($P_\phi$) components of the Poynting vector of the scattered wave are the sources of the in-plane (parallel to the scattering plane) Goos-Hänchen (GH) shift (eigenmodes- $p$ and $s$-linear polarizations) and the out of plane (transverse) Spin-Hall (SH) shift (eigenmodes- LCP/RCP) of light, respectively \cite{fedorov1955fi,gh1947,bliokh2013goos,gotte2012,imbert1972,bliokh2006prl,aiello2009,soni2014}. The exact expressions for the derived azimuthal components (which are pertinent to this study) of the SAM density ($S_\phi$) and Poynting vector ($P_\phi$) of the scattered wave are rather cumbersome, and hence not presented here. Instead, we provide below expressions for these entities in the far zone (for LCP / RCP incident wave) 
\begin{align}
S^e_\phi&=\dfrac{ \sum (2n+1) Re\{(a_n\pi_n)^*S_2\}}{\omega\rho(|S_1|^2+|S_2|^2)},   \label{eq3-6}\\
S^m_\phi&=\dfrac{ \sum (2n+1) Re\{(b_n\pi_n)^*S_1\}}{\omega\rho(|S_1|^2+|S_2|^2)}, \label{eq3-7}\\
P_\phi&=\sigma \dfrac{k|E_0|^2}{2\omega\mu\rho^3} \sin\theta \sum(2n+1) Re\{S^*_1 a_n\pi_n+S_2(b_n\pi_n)^*\},
\label{eq3-8}
\end{align}
where $\sigma=\pm1$ for RCP and LCP states, respectively, $\rho=kr$, $k$ is the wave-number in the medium and $r$ is radial distance. $\pi_n$, $\tau_n$ are angle dependent ($\theta$) functions, $S_1$ and $S_2$ are the amplitude scattering matrix elements defined as \cite{bohren}
\begin{equation}
S_1=\sum_{n=1}^\infty \dfrac{(2n+1)}{n(n+1)}(a_n\pi_n+b_n\tau_n), ~
S_2=\sum_{n=1}^\infty \dfrac{(2n+1)}{n(n+1)}(a_n\tau_n+b_n\pi_n).
\label{eq3-9}
\end{equation}
The above expressions, though approximate (valid in the far field), provide useful insights on the contributing factors to the transverse SAM and transverse spin-momentum components. For example, these indicate that both the $S_\phi$ and $P_
\phi$ components can be enhanced by the interference of scattering modes  (contain contribution of the interference term, e.g., $Re [a_2^* a_1 ]$, when only the first two lower order TM modes contribute) \cite{saha2016}. In what follows, we investigate the influence of the interference of neighboring modes on both the transverse SAM density and transverse spin-momentum components with illustrative example of resonant localized plasmon modes (TM- dipolar $a_1$ and quadrupolar $a_2$ modes) in metal nanospheres ($a<<\lambda$) and non-resonant higher order TM and TE ($a_n$ and $b_n$) modes in dielectric microspheres ($a>>\lambda$). For the sake of simplicity and ease of interpretation, we summarize the SAM density and the Poynting vector at the equatorial plane surrounding the spherical scatterer (scattering angle $\theta=\pi/2$ in Fig. \ref{fig3-1}). In the following, we cite some of the important and interesting results from Ref. \cite{saha2016} pertinent to plasmonic nanosphere and dielectric microsphere. 
\subsection*{Plasmonic sphere}
\begin{figure}[ht]
	\centering
	\includegraphics[width=0.9\linewidth]{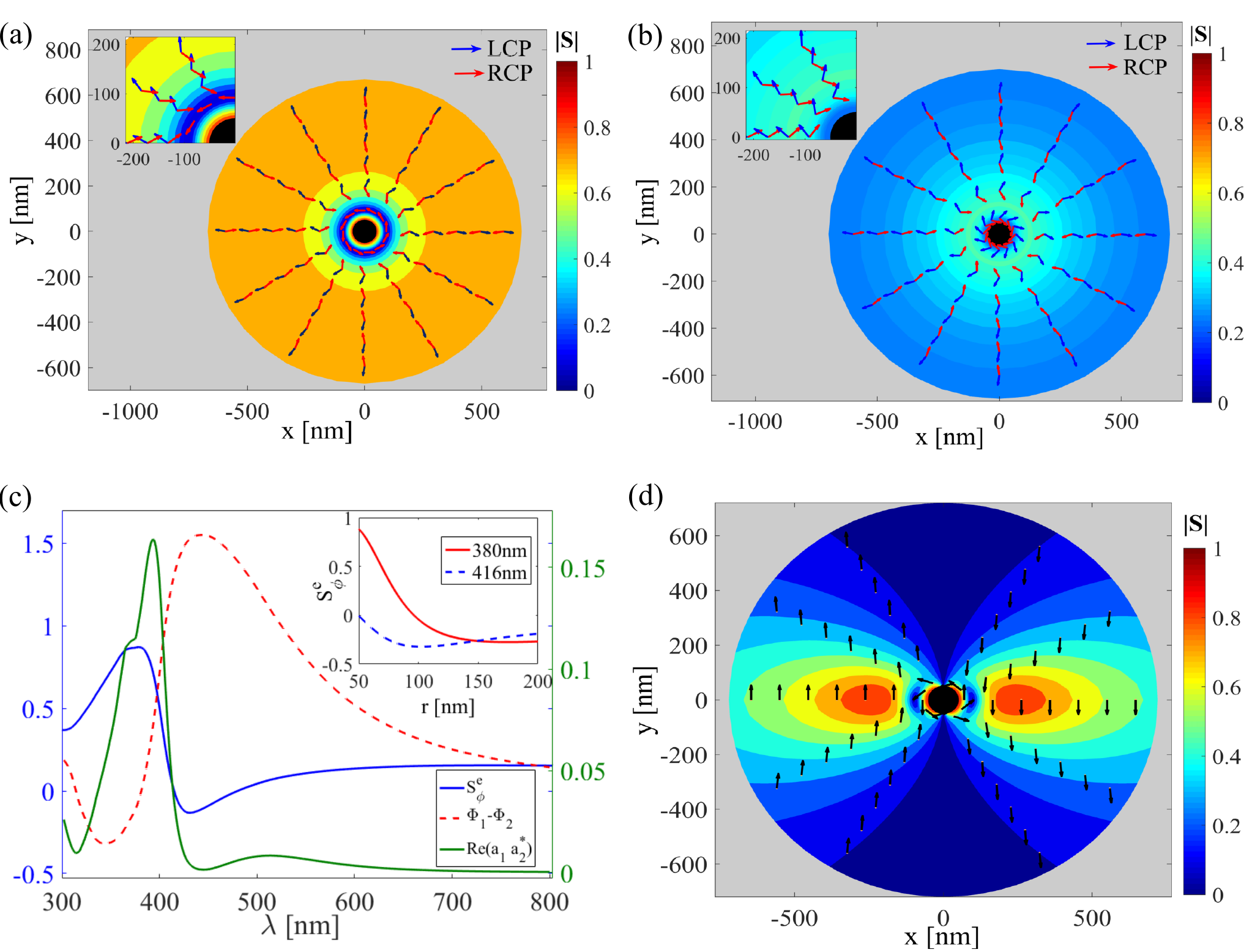}
	\caption{The computed normalized electric part of the SAM density ($\mathbf{S}^e$ shown in quiver plot) of the scattered wave from an Ag nanosphere ($a = 50$ nm) for incident LCP/RCP light, at wavelengths (a)$\lambda=380$ nm and (b)$\lambda=416$ nm; respectively.  (c) The wavelength dependence of the strength of the interference of TM dipolar $a_1$ and the quadrupolar $a_2$ plasmon modes ($Re[a_2^*a_2]$) (blue dotted line, right axis) and the phase difference ($\phi$ in rad) between them (red dashed line, left axis). The wavelength dependence of the transverse SAM density ($S_\phi^e$) at $r = 50$ nm (black solid line, left axis). The inset shows the comparison of the radial dependence of $S_\phi^e$ at 380 nm (black solid line) and 416 nm (red dotted line). (d) Quiver plot of $\mathbf{S}^e$ of the scattered wave at $\lambda=380$ nm for incident $x$- polarized light. The magnitudes of $\mathbf{S}^e$ are represented by color code and the corresponding color bars are displayed. The black circle represents the sphere at the equatorial plane here and in the subsequent figures. Adapted with permission from Ref. \cite{saha2016}, OSA Publishing.}
	\label{fig3-2}
\end{figure}
\par
In case of plasmonics sphere, our focus is on the localized plasmon resonances and we choose a sphere whose dimension is much smaller than the incident wavelength. Fig. \ref{fig3-2} shows the computed normalized SAM density ($\mathbf{S}$) of the scattered wave from a metal (silver, Ag) nanosphere of radius $a = 50$ nm with surrounding medium as water. For these calculations, the dielectric constant of Ag was taken from literature \cite{johnson1972}. Here, we have chosen to display the electric component of the SAM density ($\mathbf{S}^e$) only because for this metal nanosphere, scattering is primarily contributed by the TM- dipolar ($a_1$) and quadrupolar ($a_2$) localized plasmon modes only \cite{soni2014} (accordingly the transverse SAM $S_\phi^e$ has only electric contributions). The SAM densities of the scattered waves are shown for two wavelengths, $\lambda=380$ nm (Fig. \ref{fig3-2}a) and $\lambda=416$ nm (Fig. \ref{fig3-2}b), for incident RCP / LCP light. Several interesting observations are in place. In the near field, the SAM density is dominated by the transverse component ($S_\phi^e$), manifesting as an azimuthal structure of spin, which is nearly independent of the input LCP / RCP state (Fig. \ref{fig3-2}a and \ref{fig3-2}b). As one moves away, the conventional longitudinal component of the spin ($S_r^e$) starts dominating, and in the far field ($r>500$ nm), the spin distribution restores the usual radial structure and its expected dependence on the input circular polarization state (opposite for LCP / RCP states). These features are common for both $\lambda=380$ nm and 416 nm. However, the magnitude of the helicity-independent transverse SAM density at the near field is relatively stronger for $\lambda=380$ nm as compared to 416 nm. In order to understand this, in Fig. \ref{fig3-2}c, we have shown the wavelength dependence of the strength of the interference of the two plasmon modes ($Re[a^*_2a_1] $), and the phase difference ($\phi$) between them. The wavelength dependence of the computed transverse (azimuthal) SAM density ($S_\phi^e$, which is independent of the input LCP / RCP states) at a radial distance $r = 50$ nm is shown in the same figure. A comparison of the radial dependence of $S_\phi^e$ at 380 nm and 416 nm is also separately shown in the inset of the figure. As previously noted, the 50 nm Ag sphere exhibits prominent electric dipolar ($a_1$) and quadrupolar plasmon resonances with their peaks at $ \sim 500$ nm and 380 nm respectively \cite{soni2014,shu2009}. As evident, $S_\phi^e$ attains its maximum value at  $\lambda=380$ nm, corresponding to the spectral overlap region of the two resonances, where the strength of the interference is also maximum (phase difference between the $a_1$ and the $a_2$ modes $\phi \sim 0^\circ$ leading to constructive interference). Thus, the constructive interference of the two neighboring resonance modes leads to significant enhancement of the helicity-independent transverse SAM density in the near field ($S_\phi^e$ at 380 nm $> S_\phi^e$ at 416 nm, inset of Fig. \ref{fig3-2}c) \cite{saha2016}. This is the key result of this study. Finally, in Fig. \ref{fig3-2}d, the distribution of the SAM density ($\mathbf{S}^e$) at $\lambda=380$ nm is displayed for incident $x$-polarized light for the same Ag nanosphere. Even for incident linearly polarized light (which does not possess any spin contribution per se), there are distinct regions in the near field where the transverse SAM density assumes significant magnitudes. Clearly, the phase shifted longitudinal (radial) component of the scattered electric field is the source of this dominating transverse SAM component ($S_\phi^e$) of the scattered wave for linear polarization excitation.  These results clearly demonstrate that the observed transverse SAM density of the scattered wave in the near field region is independent of the input helicity and polarization state of light and is similar in spirit to those observed for the evanescent waves \cite{bliokh2014}.
\par
\begin{figure}[ht]
	\centering
	\includegraphics[width=0.9\linewidth]{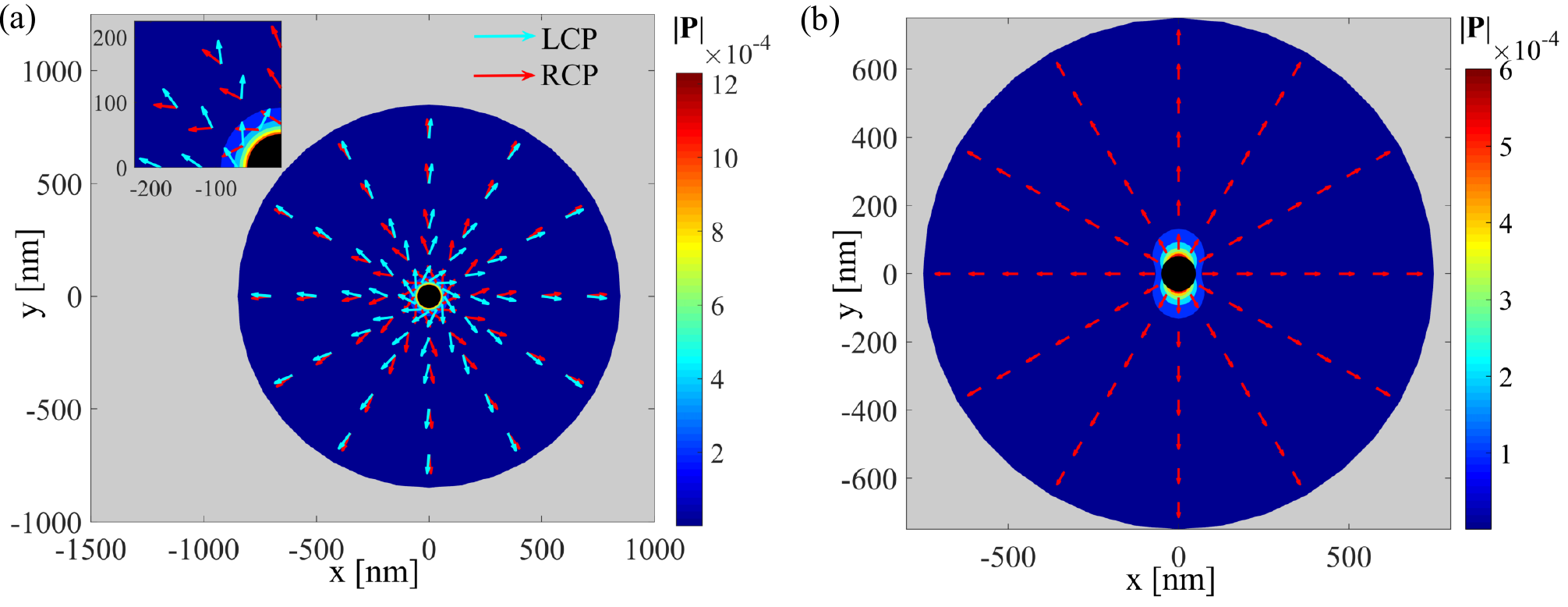}
	\caption{The computed Poynting vector $\mathbf{P}$ distribution of the scattered wave ($\lambda=380$ nm) from the Ag nanosphere corresponding to Fig. \ref{fig3-2}, for incident (a) LCP / RCP light and (b) $x$–linearly  polarized light. The magnitudes of $\mathbf{P}$ are represented by color code and the corresponding color bars are displayed. Adapted with permission from Ref. \cite{saha2016}, OSA Publishing.}
	\label{fig3-3}
\end{figure}

In Fig. \ref{fig3-3}, we show the other important entity, the Poynting vector ($\mathbf{P}$) distribution of the scattered wave ($\lambda=380$ nm) from the same Ag nanosphere. In Fig. \ref{fig3-3}a, we have shown the computed Poynting vector distribution at the equatorial plane of the sphere for input RCP / LCP. It is observed that at the near field, the flow of Poynting vector is dominated by the transverse component which is opposite for input RCP / LCP; and with increasing distance, longitudinal component of $\mathbf{P}$ takes over and usual polarization independent flow of $\mathbf{P}$ is restored. In Fig. \ref{fig3-3}b, the Poynting vector flow is shown for $x$-polarized input light. From the figure, we see that the flow of Poynting vector is radial ( both for near and far field) in this case. Thus the azimuthal component of $\mathbf{P}$ can be identified as the spin-dependent transverse (spin) momentum of Poynting vector \cite{saha2016}.
\par
Thus results show that the interference of lower order TM modes in plasmonic metal nanoparticles leads to enhancement of the transverse SAM and momentum densities. It is therefore interesting to study how these quantities are affected when several higher order TM and TE ($a_n$ and $b_n$) modes are simultaneously excited. For this purpose, we choose a non-resonant dielectric microsphere ($a =  1$ $\mu$m and refractive index=1.6) with surrounding medium as air.
\subsection*{Dielectric microsphere}
\begin{figure}[ht]
	\centering
	\includegraphics[width=\linewidth]{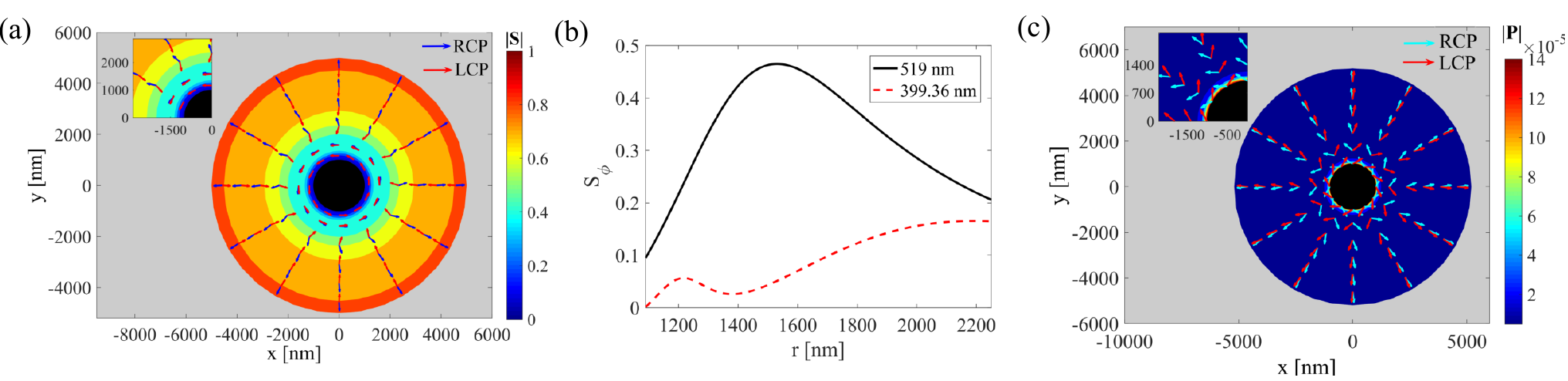}
	\caption{The computed SAM density ($\mathbf{S}$) and Poynting vector ($\mathbf{P}$) distributions of the scattered wave from a dielectric microsphere ($a = 1$ $\mu$m and refractive index$=1.6$). (a) Quiver plot of $\mathbf{S}$ for $\lambda=519$  nm.  (b) Comparison of the radial dependence of transverse (azimuthal) SAM density ($S_\phi$) at $\lambda=519$ nm (non-resonant scattering system, black solid line) and $\lambda=399.36$ nm (resonant scattering system, red dotted line). (c) Quiver plot of $\mathbf{P}$ of the scattered wave $\lambda=519$ nm for incident LCP/RCP light. Adapted with permission from Ref. \cite{saha2016}, OSA Publishing.}
	\label{fig3-4}
\end{figure}

The computed normalized SAM density ($\mathbf{S}$) and Poynting vector ($\mathbf{P}$) distributions of the scattered wave from this dielectric scatterer are summarized in Fig. \ref{fig3-4}.  We choose two different wavelengths, $\lambda=519$ nm and $\lambda=399.36$ nm. At $\lambda=519$ nm, the dielectric microsphere acts as a non-resonant scatterer, where scattering is contributed simultaneously by several higher order TM and TE modes. At $\lambda=399.36$ nm, on the other hand, scattering is primarily contributed by the resonant TM $a_{20}$ mode. The computed SAM density ($\mathbf{S}=\mathbf{S}^e+\mathbf{S}^m$, shown in Fig. \ref{fig3-4}a) at $\lambda=519$ nm exhibit similar trends as previously observed – (a) the SAM density is dominated by the transverse spin ($S_\phi$) in the near field (exhibiting azimuthal structure), which is independent of the incident LCP/RCP state; and (b) in the far field, the spin distribution restores the usual radial structure and is also opposite for incident LCP / RCP states \cite{saha2016}. 
\par
Importantly, the magnitude of the helicity-independent transverse spin ($S_\phi$) and the corresponding spatial extent over which it dominates is much larger for the non-resonant dielectric scatterer ($r\sim2000$ nm) as compared to that observed for the plasmonic Ag nanosphere ($r\sim150$ nm). Note that unlike the plasmonic scatterer, for this non-resonant dielectric scatterer, $S_\phi$ has both electric and magnetic contributions ($S_\phi^m\neq0$) (originating from non-zero longitudinal components of both the scattered electric and the magnetic fields). Even though, the resonant dielectric scattering system (same scatterer but at $\lambda=399.36$ nm) also exhibits similar helicity-independent transverse SAM, its magnitude is considerably weaker. This is shown in Fig. \ref{fig3-4}b, where a comparison of the radial dependence of the transverse SAM density ($S_\phi$) at $\lambda=519$ nm and $\lambda=399.36$ nm is displayed.   Thus the observed higher magnitude and spatial extent  of polarization-independent transverse SAM density for the non-resonant dielectric scatterer compared to both the resonant dielectric scattering system and the plasmonic scatterer implies that simultaneous presence of  higher order TM and TE scattering modes and their interference is responsible for enhancement of the transverse spin. Finally, in Fig. \ref{fig3-4}c, we show the Poynting vector ($\mathbf{P}$) distribution of the scattered wave ($\lambda=519$ nm) for incident LCP / RCP light, from the same dielectric microsphere system. Once again, input circular polarization-dependent transversal energy flow in the near field is evident, with its spatial extent ($r\sim1500$ nm) being considerably larger as compared to that observed for the plasmon resonant Ag nanosphere ($r\sim150$ nm) \cite{saha2016}. 

\section{Focused radially and azimuthally polarized vector beams: Purely transverse SAM and longitudinal momentum densities \label{sec3-2}}
As shown in Section \ref{sec3-1} the scattering of plane waves from micro and nano optical systems can lead to helicity independent transverse SAM density. We discussed that the SAM density of the scattered wave is dominated by the transverse component mainly in the near field and in the far field, conventional longitudinal component takes over. In a recent study we reported that using focused radially and azimuthally polarized input vector beams, one can observe a purely transverse SAM density with a longitudinal Poynting vector in field scattered from the spherical particle \cite{pra2018}.
\par
The field scattered from a complex source radially ($E_s^{rad}/H_s^{rad}$) and azimuthally ($E_s^{azi}/H_s^{azi}$) polarized focused beam can be written as \cite{orlov2010}:
\begin{eqnarray}
\mathbf{E}_s^{rad}=&- \sum\limits_{n=1}^\infty a_n \alpha_n {\mathbf{N}}_{e0n},~~~~ \mathbf{H}_s^{rad}=&\dfrac{ik}{\omega \mu} \sum\limits_{n=1}^\infty a_n \alpha_n {\mathbf{M}}_{e0n}, \label{eq3-10} \\ \mathbf{E}_s^{azi}=&- \sum\limits_{n=1}^\infty b_n \alpha_n {\mathbf{M}}_{e0n},~~~~ \mathbf{H}_s^{azi}=&\dfrac{ik}{\omega \mu} \sum\limits_{n=1}^\infty b_n \alpha_n {\mathbf{N}}_{e0n}. \label{eq3-11}
\end{eqnarray}
Here, $\alpha_n$ determines the amplitude of the scattered field depending upon focusing of the incident beam and it is expressed as \cite{orlov2010,pra2018}:
\begin{equation}
\alpha_n=\dfrac{kz_0(2n+1)}{sinh(kz_0)}j_n(ikz_0),
\label{eq3-12}
\end{equation}
where $kz_o$ is the collimation distance of the focused beam with beam waist $w_o$ and Rayleigh distance $z_o=kw_o^2/2$.
\begin{figure}[ht]
	\centering
	\includegraphics[width=0.9\linewidth]{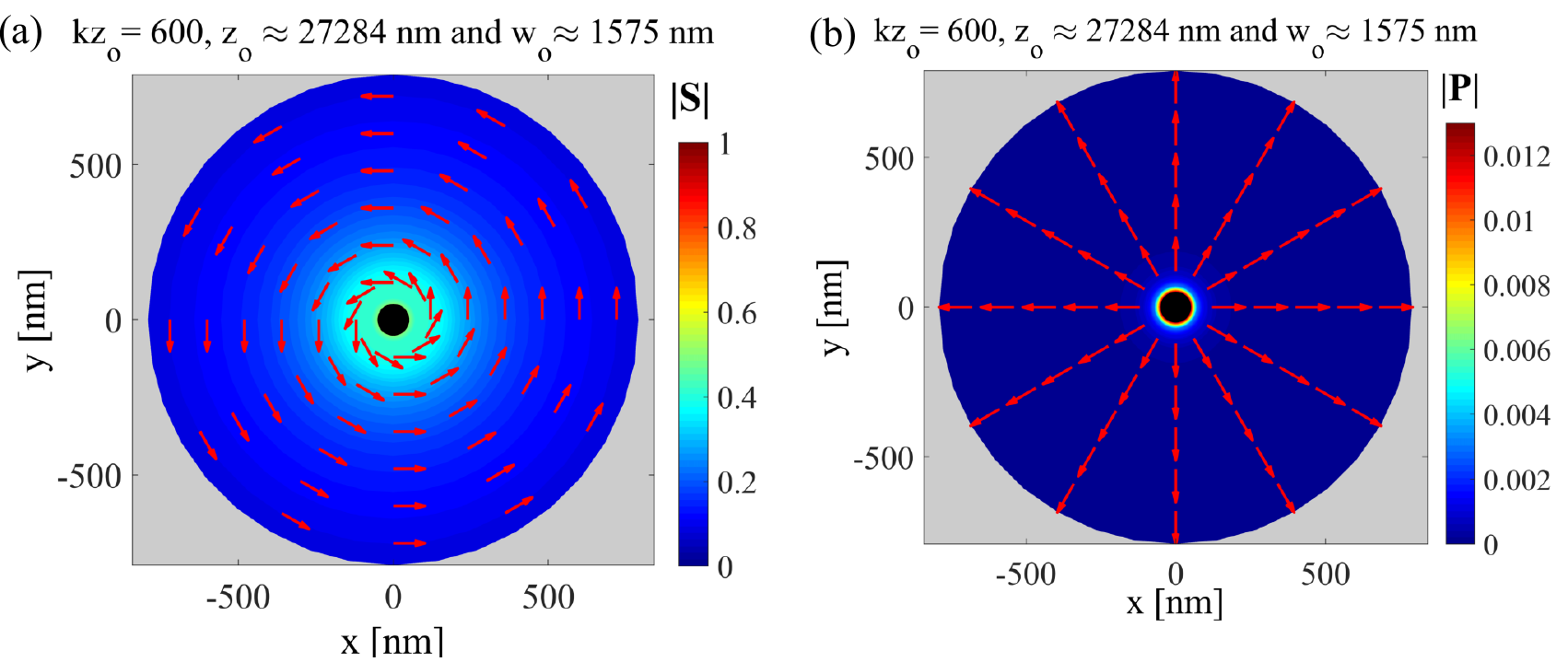}
	\caption{The spatial distribution of the computed (a) normalized SAM $\mathbf{S}$ density and (b) Poynting $\mathbf{P}$ vector of the scattered wave is shown in the equatorial plane of a spherical plasmonic silver (Ag) scatterer ($a=50$ nm) kept in water, with incident focused radially polarized beam with Rayleigh distance $z_0 \approx 27284$ nm beam spot $w_o\approx1575$. Adapted with permission from Ref. \cite{pra2018}.}
	\label{fig3-5}
\end{figure}
It can be observed from Eqs. (\ref{eq3-10})-(\ref{eq3-11}) that for a given input radially or azimuthally polarized beam either TM or TE ($a_n$ or $b_n$) modes will contribute to the scattered field of the particle. Thus for incident radial polarization, TM ($a_n$) modes will contribute to the SAM density of scattered field; while for incident azimuthal polarization, SAM density will have contribution from TE ($b_n$) modes only. It is to be noted that in such a case the SAM density is purely transverse in nature. These observations are independent of the size of sphere or the wavelength of light incident.
\par
The spatial distribution of the calculated normalized SAM density and Poynting vector of the scattered field for an incident focused radially polarized beam of wavelength $\lambda=380$ nm on the same Ag nanosphere (discussed in section \ref{sec3-1}) is shown in Fig. \ref{fig3-5}.  The wavelength is chosen such that the contribution from the effective electric dipolar ($\alpha_1a_1$) and electric quadrupolar ($\alpha_2a_2$) modes interfere constructively as in case of plane waves. The Fig. \ref{fig3-5}a and Fig. \ref{fig3-5}b clearly demonstrate the completely transverse nature of SAM density and the longitudinal nature of Poynting vector. It is to be noted from Eq. (\ref{eq3-12}) that the amplitude parameter $\alpha_n$ depends on collimation distance ($kz_o$) of the input beam. Thus, the SAM density and Poynting vector of the scattered field can also be tuned by changing the focusing parameter of the input focused beam on the spherical particle.
\section{Layered spherical scatterer: Effect of mode-coupling on the transverse SAM density  \label{sec3-3}}
\begin{figure}[ht]
	\centering
	\includegraphics[width=0.7\linewidth]{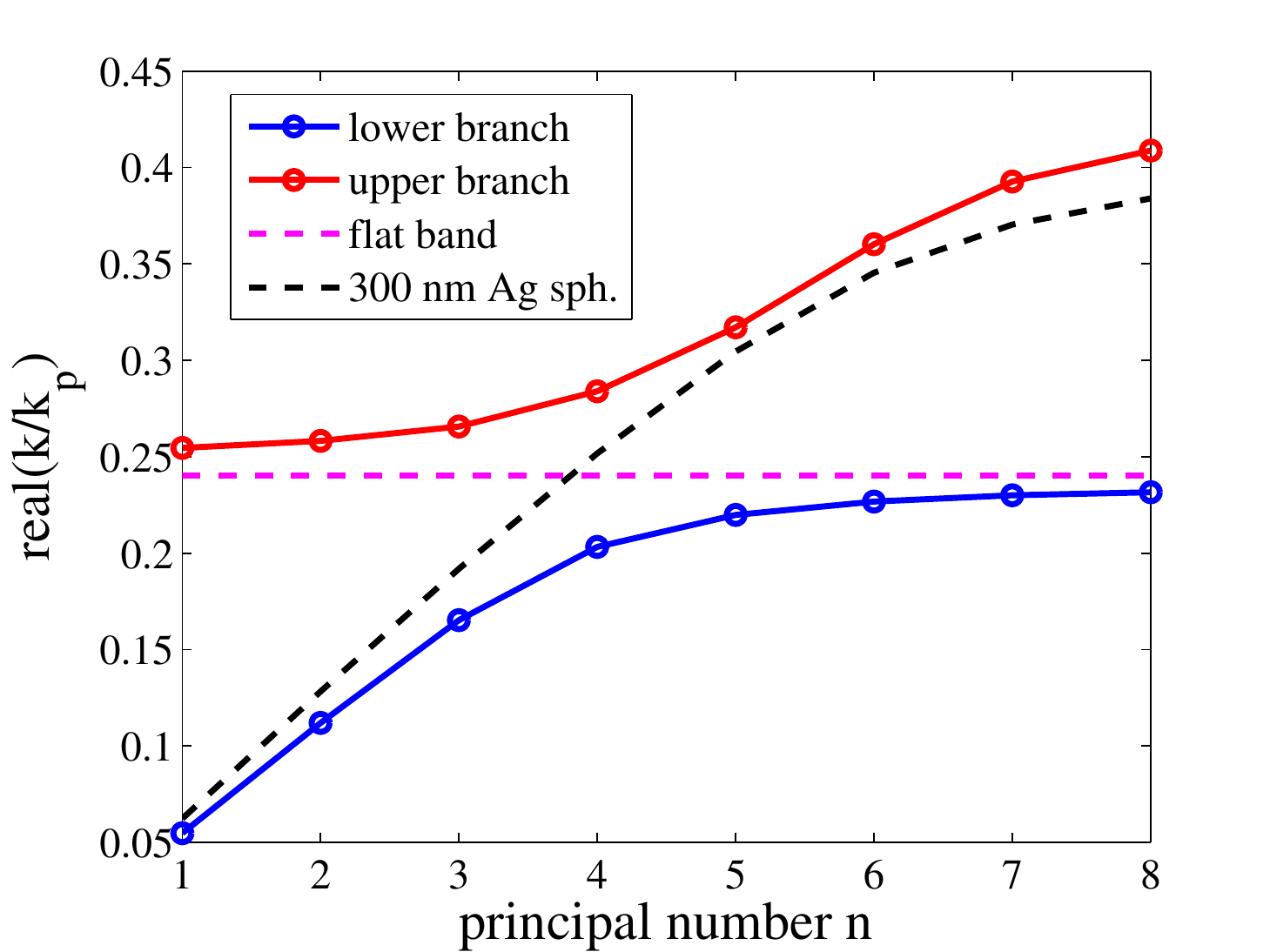}
	\caption{The dispersion diagram of Ag-Si-Ag sphere with $R=300$ nm, $L=30$ nm and $T=10$    nm. The upper and lower branches of the dispersion diagram are shown by red and blue curves. The flat back is shown by black  color dashed line and the flat band of the dispersion curve is represented by pink dashed line. Republished with permission of IOP Publishing, from Effects of mode mixing and avoided crossings on the transverse spin in a metal-dielectic-metal sphere, S. Saha, A. K. Singh,, N. Ghosh, and S. D Gupta, J. Opt., $\mathbf{20}$ (2), 025402, 2018; permission conveyed through Copyright Clearance Center, Inc.}
	\label{fig3-6}
\end{figure}
In this section, we review the effect of mode mixing and avoid crossing on the transverse SAM density of light \cite{saha2018}. We consider the specific case of a layered sphere comprising of metal core of radius $R$, a dielectric layer of thickness $L$ and an outer metal layer boundary of thickness $T$, thus the total radius of the sphere is $W=R+L+T$. Now the dispersion relation for such a layered sphere can be obtained by solving the following determinant equation \cite{rohde2007}: 
\begin{equation}
\label{eq3-13}
\begin{vmatrix}
|u| & |A| \\|B| & |V|
\end{vmatrix}=0,
\end{equation}
where |$u|=0$ gives the eigen-modes of a metal sphere of radius $W$ placed in a dielectric surrounding medium of permittivity $\epsilon_0$; $|V|=0$ gives the eigen-modes of a MDM sphere of radius $S=R+L$ with infinite thick metal surrounding; the off-diagonal elements $|A|$ and $|B|$ are the coupling terms.
\par
\begin{figure}[ht]
	\centering
	\includegraphics[width=0.9\linewidth]{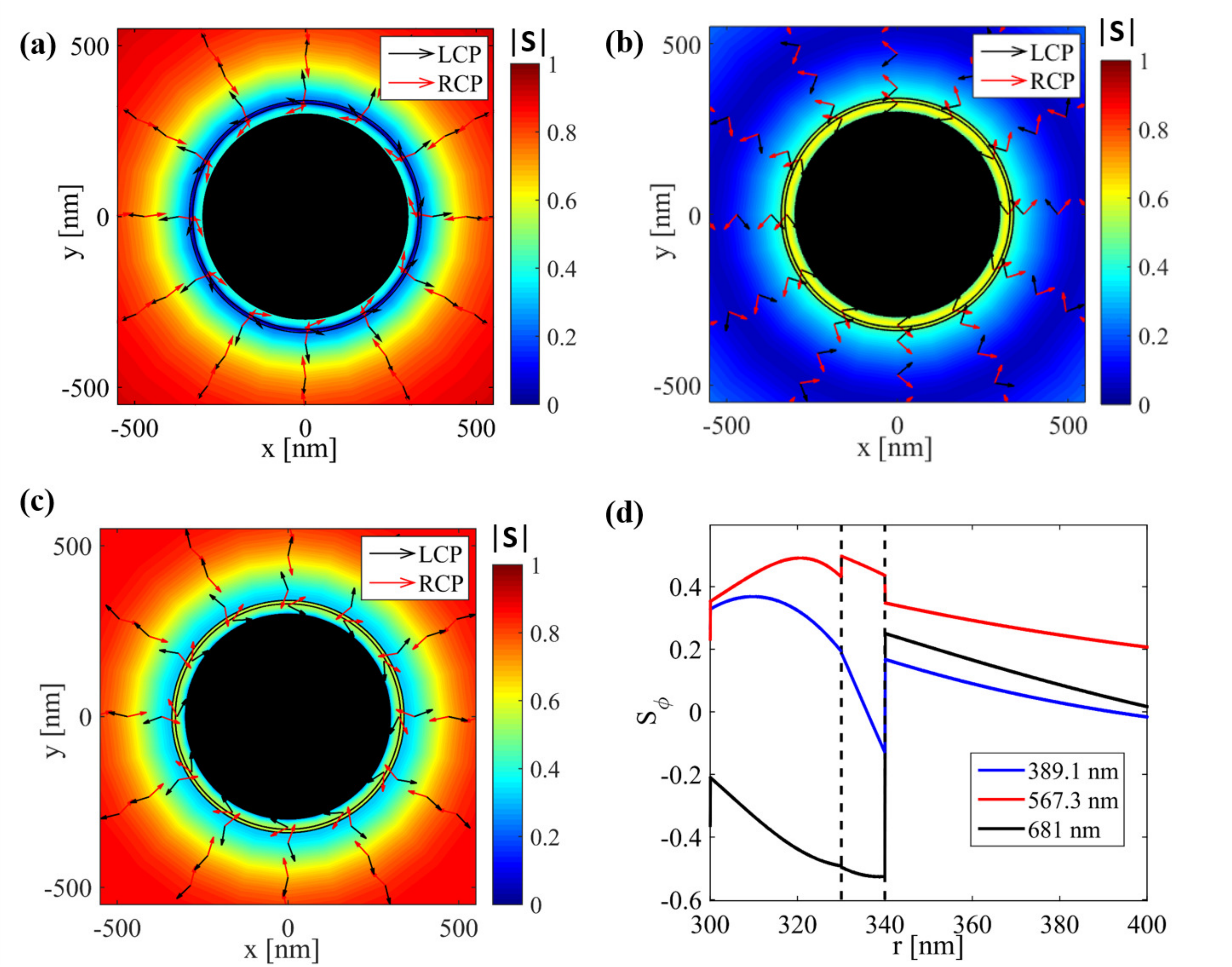}
	\caption{The computed normalized SAM density $\mathbf{S}$ at the equatorial plane of the Ag-Si-Ag sphere for input LCP/RCP is shown atwavelength (a) 389.1 nm (below the avoided crossing), (b) 567.3 nm (at the avoided crossing) and (c) 681 nm (above the avoided crossing); (d) the radial variation of the transverse SAM density ($S_\phi$) at the three wavelengths. Republished with permission of IOP Publishing, from Effects of mode mixing and avoided crossings on the transverse spin in a metal-dielectic-metal sphere, S. Saha, A. K. Singh,, N. Ghosh, and S. D Gupta, J. Opt., $\mathbf{20}$ (2), 025402, 2018; permission conveyed through Copyright Clearance Center, Inc.}
	\label{fig3-7}
\end{figure}
In our case, we choose a silver (Ag) core of radius $R=300$ nm, a silicon (Si) layer of thickness $L=30$ nm and a Ag outer layer of thickness $T=10$ nm. The dispersion diagram for this MDM sphere is shown in Fig. \ref{fig3-6}. Thus from the dispersion diagram, we see a well-defined avoided crossing near the principal mode number $n=4$. The flat band in the dispersion diagram corresponds to the surface plasmon frequency $k_{sp}/k_p =0.24$ (here $k_p=\frac{\omega_p}{c}$, $\omega_p$ is the plasma frequency)  \cite{rohde2007,saha2018}. The avoided crossing in this case results from the coupling between the Mie modes of the metal core with the surface plasmons at the exterior surface of the dielectric shell \cite{saha2018}. 
\begin{figure}[ht]
	\centering
	\includegraphics[width=\linewidth]{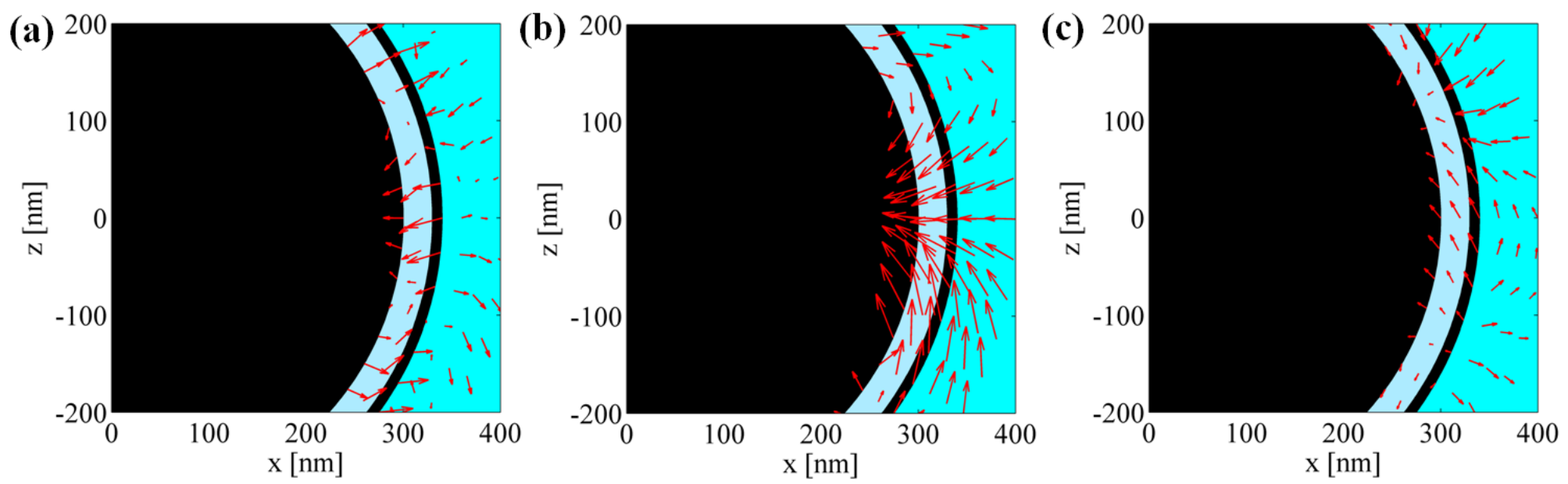}
	\caption{Circulation of the field in the $x-z$ plane of the sphere for (a) 389.1 nm (b) 567.3 nm and (c) 681 nm. Republished with permission of IOP Publishing, from Effects of mode mixing and avoided crossings on the transverse spin in a metal-dielectic-metal sphere, S. Saha, A. K. Singh,, N. Ghosh, and S. D Gupta, J. Opt., $\mathbf{20}$ (2), 025402, 2018; permission conveyed through Copyright Clearance Center, Inc.}
	\label{fig3-8}
\end{figure}
\par
We now discuss the structure of the SAM density across the avoided crossing of the MDM sphere. For this purpose, we choose three wavelengths 389.1 nm (below the avoided crossing), 567.3 nm (at the avoided crossing) and 681 nm (above the avoided crossing). Fig.  \ref{fig3-7}a-c show the computed normalized SAM density at the equatorial plane of the sphere at 389.1 nm, 567.3 nm and 681  nm, respectively.  From these figures, it is observed that at the crossing the helicity-independent transverse component of the SAM density of the scattered field is dominant over the longitudinal component (helicity dependent) of SAM density; while at wavelengths away from the crossing, the contribution of the azimuthal component is significantly reduced. In Fig. \ref{fig3-7}d, the radial variation of the transverse SAM density of the scattered wave is shown. Thus we clearly see higher transverse SAM density at the avoided crossing compared to wavelengths away from the crossing. Note that the enhancement of the transverse component of the SAM density at $\lambda=567.3$ nm at the equatorial plane resulted from the fact that the field becomes highly rotational at the crossing compared to the same away from the crossing (Fig. \ref{fig3-8}).  
\par
In a nutshell, we have demonstrated the presence of  helicity-independent transverse SAM and helicity-dependent (spin) momentum components in case of scattering of plane waves and focused vector beams from a spherical scatterer. These extraordinary properties of the fields are usually associated with evanescent fields. Our study reveals that these properties of the field are generic in any scattering problem and can also be realized from a simple scatterers e.g. a plasmonic nano-sphere or a dielectric microsphere, MDM sphere etc.

\chapter{Experimental studies of transverse spin and momentum densities of light \label{chapter4}}

In this chapter, we review some of the remarkable experiments on the detection and measurement of transverse SAM and momentum densities of structured optical fields. These transverse spin and momentum densities are observed in case of two-wave interference \cite{bekshaev2015}, evanescent field \cite{bliokh2014}, surface plasmon polaritons \cite{bliokh2012} and whispering gallery mode resonators \cite{junge2013} etc. The linear momentum density can be decomposed into orbital and spin parts. As noted earlier, the spin part of the momentum density, known as the Belinfante's spin-momentum density, does not exert optical force on an absorbing Rayleigh particle (radius$<<$wavelength), but it can interact with a Mie particle for which radius is comparable to the wavelength. This feature of the spin-momentum makes it challenging to detect it experimentally. Hence it is extremely useful to investigate the transverse spin and Belinfante's spin momentum density experimentally both for fundamental interests and potential applications.
\par 
To the best of our knowledge, the first experimental detection of the electric part of transverse spin density of a tightly focused linear and radially polarized beam was done by Neugebauer {\textit{et al.}}  using a dipole like plasmonic probe particle \cite{measuring2015}. Antognozzi {\textit{et al.}} reported the first direct measurement of the polarization dependent optical momentum and transverse spin-dependent force using a nano-cantilever immersed in an evanescent field \cite{natphys_2016}. Recently, Yang {\textit{et al.}} \cite{nanoscale2018} reported the coupling between transverse spin of a focused field with that of a surface plasmon polaritons experimentally. Such coupling is shown to have useful applications for particle-identification and field mapping. In another recent experimental study, Garoli {\textit{et al.}} showed the coupling of transverse SAM of plasmons propagating on a nanotaper to the light's polarization\cite{garoli2017}. In the following, we briefly summarize the main findings of these studies.
\section{Measuring transverse SAM density of light}

\begin{figure}[ht]
	\centering
	\includegraphics[width=0.7\linewidth]{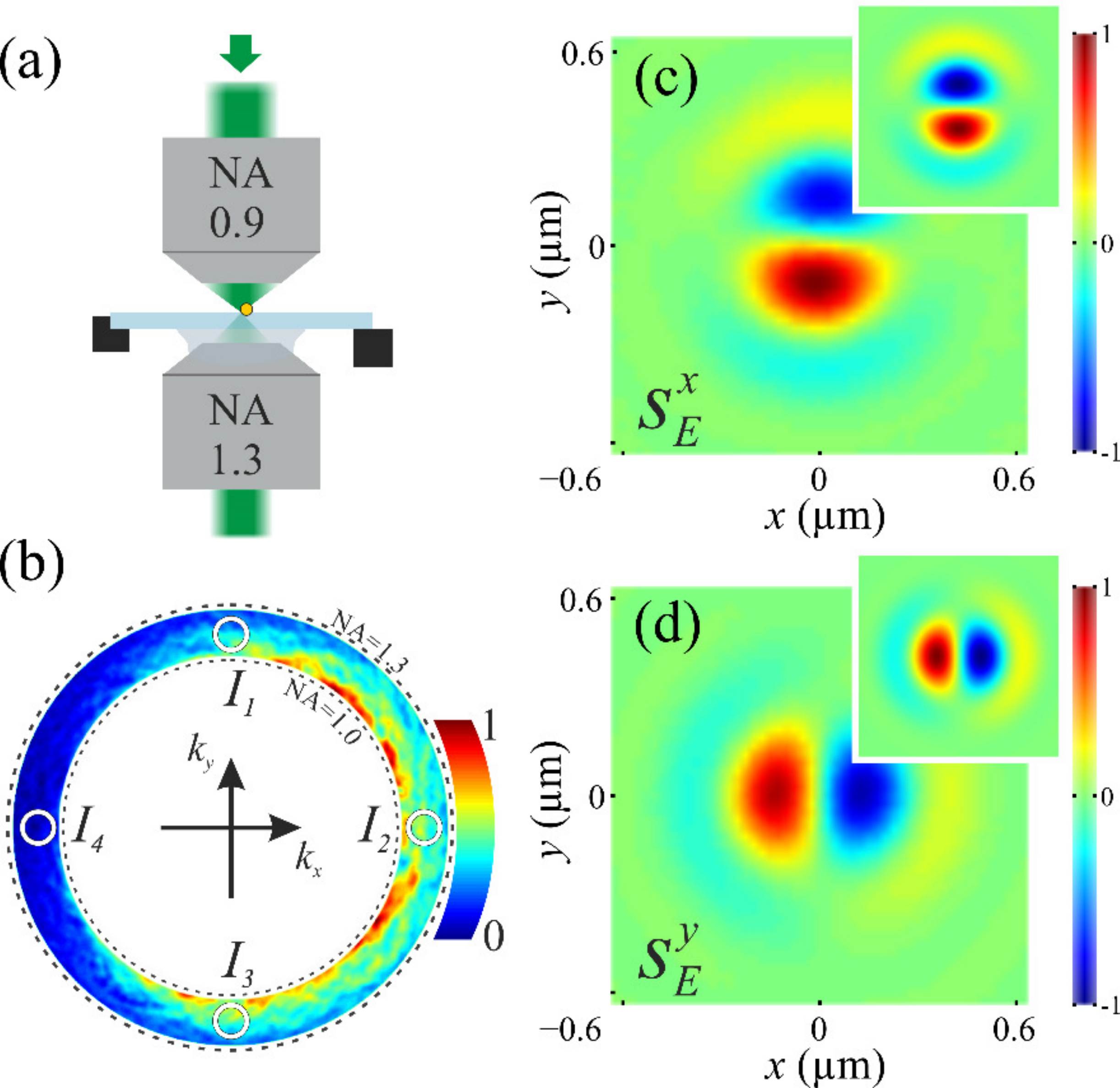}
	\caption{(a) The schematic of experimental setup. The incoming light can be of linear or radially polarized is focused on a dipolar plasmonic gold sphere kept on high dielectric constant substrate kept on a movable stage. The scattered light is then collected through a high NA objective. b) Light scattered is detected in the image plane, the white circle indicate the coordinates mentioned in the equation. The measured distribution of the components of transverse spin are shown in c-d for incident radial polarization. Inset of c and d show the corresponding calculated transverse SAM density. Here the SAM density was plotted by normalizing it by the maximum value. Adapted from Ref. \cite{measuring2015}. }
	\label{fig4-1}
\end{figure}

Neugebauer {\textit{et al.}} reported measurement of the distribution of transverse spin density components  of focused radially and linearly polarized beams \cite{measuring2015}. The idea of detection is based on near-field interference of a transversely spinning dipole. Fig. \ref{fig4-1}a shows the schematic of their experimental system where a dipolar plasmonic gold sphere is kept on dielectric-air interface under an imaging microscope. A dipole kept in close proximity of dielectric medium of high dielectric constant converts the evanescent field into propagating waves that can be observed in the far field. From the basic definition of electric spin density under the assumption of dipolar nature of scattering, the polarizibility of the scatterer depends only on the local electric field and  it was shown that \cite{measuring2015}:
\begin{eqnarray}
\Delta I^x&=&I(0,-k_y^\prime)-I(0,k_y)\propto S_E^x, \\
\Delta I^y&=&I(k_x^\prime,0)-I(-k_x^\prime,0)\propto S_E^y.
\end{eqnarray}
Here, $I(k_x^\prime,k_y^\prime)$ denotes the intensity at the coordinate ($k_x^\prime,k_y^\prime$) in the imaging plane and $S_E^{(x/y)}$ denotes the X/Y component of transverse spin.
\par 
Using the above experimental scheme, the components of transverse spin density were determined for a given position of the spherical probe particle. The particle was moved throughout the cross section of the beam to obtain the distribution of the components of transverse spin across the beam cross section. Fig.\ref{fig4-1}b shows the light scattered  in the image plane and Fig. \ref{fig4-1}c-d show the transverse SAM density components for input radially polarized beam.

\section{Direct measurements of the extraordinary optical momentum and transverse spin-dependent force using a nano-cantilever}
\begin{figure}[ht]
	\centering
	\includegraphics[width=0.7\linewidth]{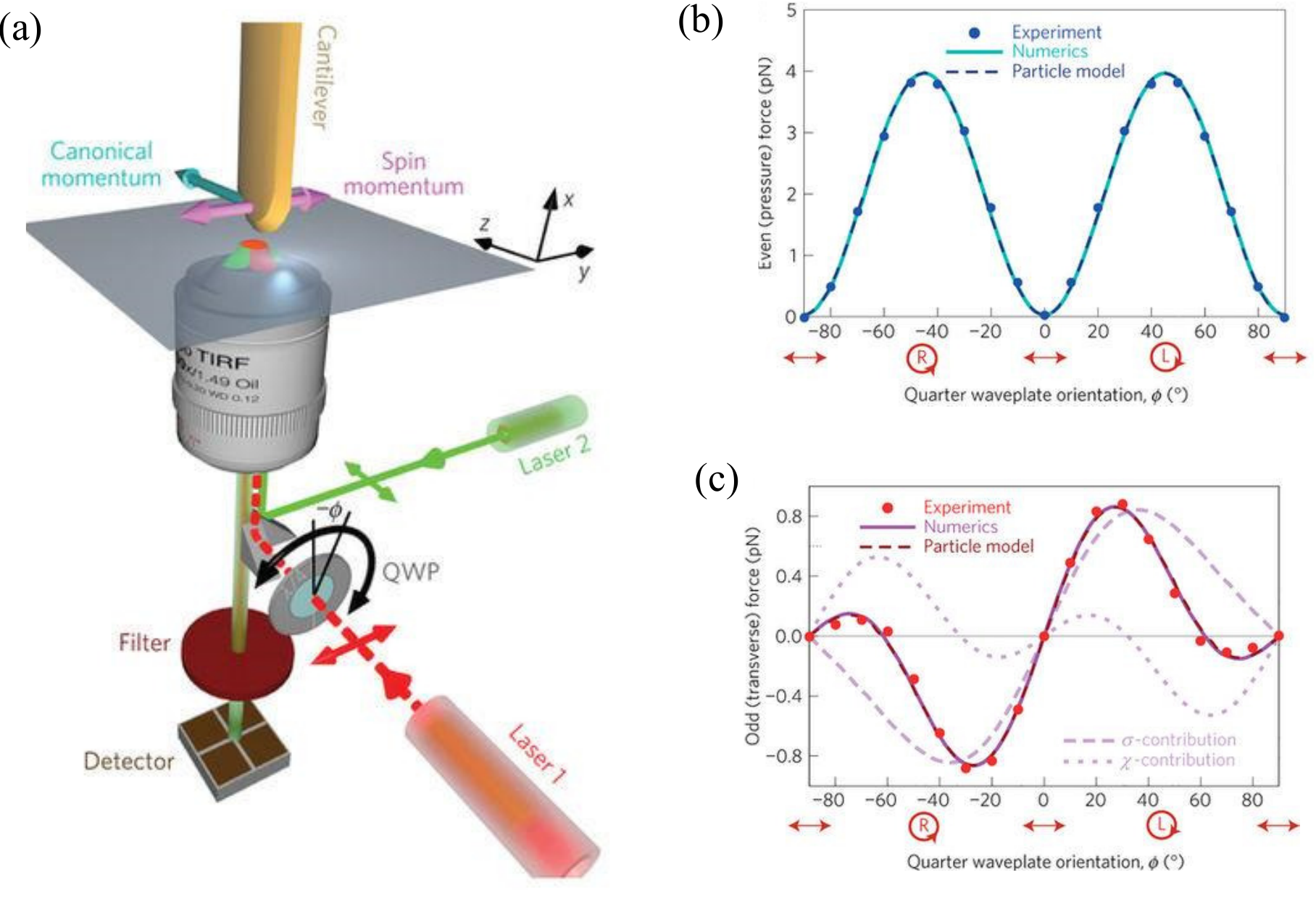}
	\caption{(a) The lateral molecular force microscope set-up: red laser 1 produces an evanescent field to be probed. Its intensity is modulated and the polarization is controlled by a rotating quarter waveplate (QWP). The green laser 2 images the position of the cantilever probing the evanescent field of laser 1; (b) Longitudinal (c) and transverse optical forces on the cantilever in the evanescent wave. Reprinted by permission from Springer Nature: Nature Physics, Direct measurements of the extraordinary optical momentum and transverse spin-dependent force using a nano-cantilever, M. Antognozzi {\textit{et al.}}, (2016).}
	\label{fig4-2}
\end{figure}
The first experimental detection of the unusual transverse momentum has been done using a femto newton-resolution nano-cantilever immersed in an evanescent optical field above the total internal reflecting glass surface \cite{natphys_2016}. The experimental set-up uses a lateral molecular force microscope (LMFM).  A red laser (laser 1) of wavelength 660 nm is used to generate $z$-propagating and $x$-decaying evanescent field at the glass-air interface using an objective based on total internal reflection system. A quarter wave plate (QWP) is used to control the polarization of the incident laser by varying the orientation angle of the QWP. The cantilever (spring constant $\gamma=2.1 \times 10^{-5}$, refractive index $n=2.3$, thickness $d=100$ nm, width $w=1000$ nm, length $l=120$ $\mu$m) is mounted vertically in the evanescent field with its tip being 30 nm above the glass coverslip and very sensitive to the transverse force in the evanescent field. The deflection of the nano-cantilever is caused due to the optical forces acting on it and recorded using a non-interferometric scattered evanescent wave (SEW) method which uses another green laser (laser 2). The reactive-ion etching in the cantilever fabrication resulting in imperfect asymmetric shape made the cantilever sensitive to the longitudinal radiation force. Thus by varying the polarization of the incident beam, both the longitudinal and transverse forces were measured. The result shows that the longitudinal force ($\sim pN$) is always the same for input RCP/LCP (Fig. \ref{fig4-2} b). There is also another transverse force acting on the cantilever which changes its sign with the sign of polarization and it is two-orders of magnitude smaller than the longitudinal force (i.e. the radiation force).  This transverse force thus bears the signature of the Belinfante spin momentum part of the Poynting vector. 
\section{Optical transverse spin coupling through a plasmonics nanoparticle }

As already discussed in Chapter \ref{chapter1} that focused beam possesses a transverse spin component which arises due to a phase-difference of $\pi/2$ between the longitudinal and transverse field components. In a recent experiment, Yang {\textit{et al.}} have reported that the transverse spin of a focused beam can be coupled with that of a surface plasmon polariton (SPP) \cite{nanoscale2018}. The study showed that the coupling between the two may lead to useful application in the particle-identification and field-mapping. In the experiment, a metal nanoparticle (Ag / Au) on a metal film (Ag) is used to couple the local SAM of a focused beam with that of a SPP. It is shown that the coupling is highly sensitive to the material and  size of the nanoparticle, distance from the surface of the metal film and incident polarization. The coupling is controlled by the strong-coulomb interaction between the metal particle and the film. The results of the study indicate that a focused beam can be employed to characterize the nanoparticle or vice versa. In the following, we briefly summarize some of the important results of their study and direct the interested readers to the original paper for more detailed discussion \cite{nanoscale2018}.
\begin{figure}[ht]
	\centering
	\includegraphics[width=0.9\linewidth]{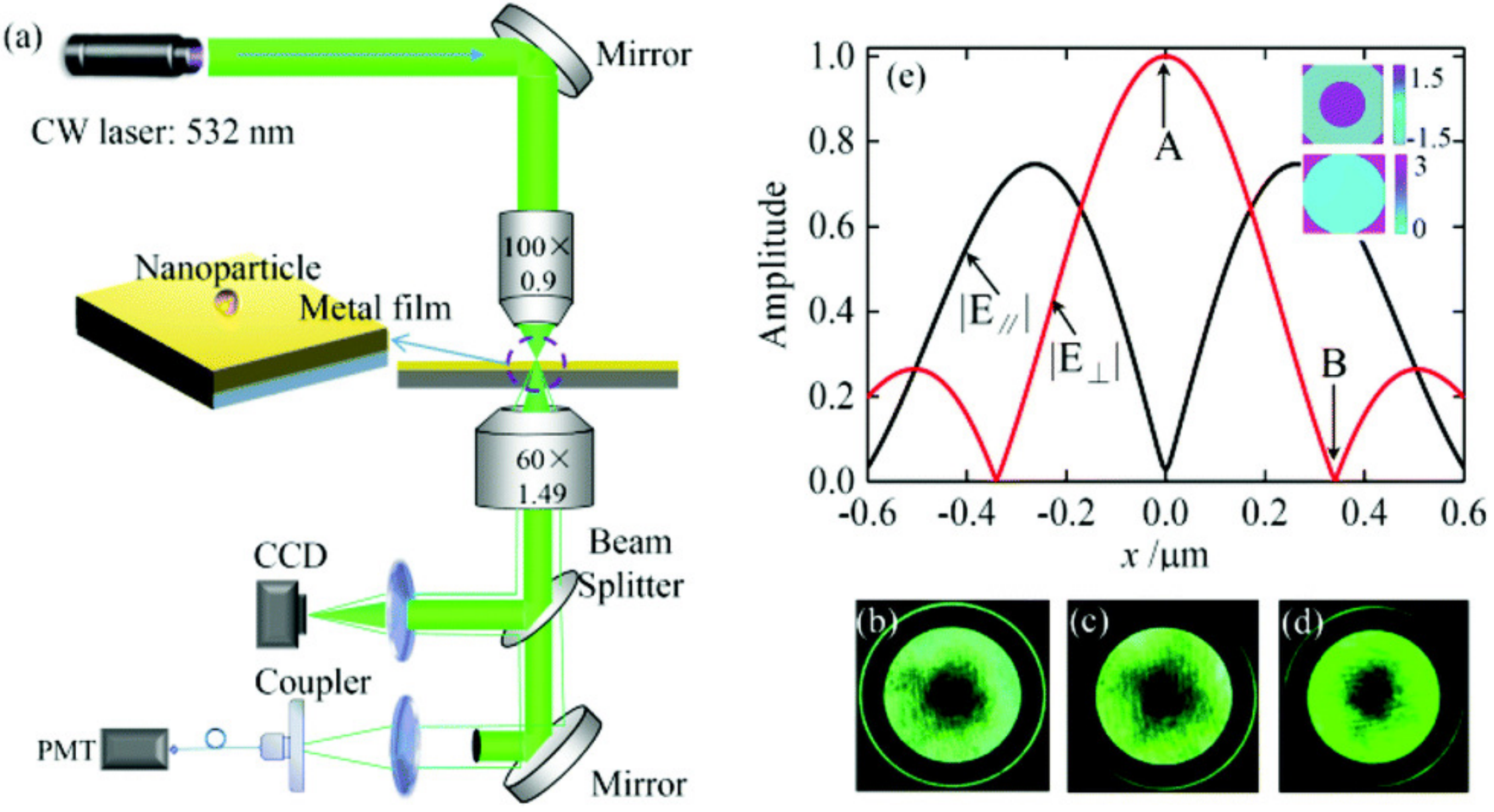}	
	\caption{(a) Schematic of the experimental set up for coupling the transverse SAM of the focused field and a SPP via a plasmonic nanoparticle. (b-d) SPCE images recorded at the CCD camera at the back focal plane with an Ag nanoparticle for three different locations of the beam. (e) theoretically calculated longitudinal $(E_\perp)$ and transverse $(E_\parallel)$ components of the electric field at the focal plane of a tightly focused radially polarized input beam. Inset shows their phase delay. Adapted from Ref. \cite{nanoscale2018} with permission of The Royal Society of Chemistry.}
	\label{fig4-3}
\end{figure}
\par
Fig. \ref{fig4-3}a shows the experimental set-up which uses a 532 nm laser source coupled to the sample using an objective lens ($NA=0.9, 100\times$) to tightly focus the beam on the sample and the scattered light is then collected using another high NA oil immersion objective lens ($NA=1.43, 60\times$). The surface plasmon coupled emission (termed as SPCE) is recorded using a CCD camera at the back focal plane of the objective and the intensity is recorded using a photomultiplier (PMT) at the imaging plane. To extract the SPCE signal from the entire beam an opaque circular disk is employed at the PMT-branch to cut-off the central transmitted light. The SPCE signal is then focused and coupled to PMT using fiber coupler and mutlimode fibre to record the SPCE signal emitted from the nanoparticle.
\par 
The sample comprises of three layers: an air substrate / silver film / silica substrate with few immobilized Ag and Au nanoparticles distributed at the air / silver interface. While interacting with the focused field, the nanoparticle scatters light (Rayleight scattering) towards the substrate side via SPCE. A monolayer of 4-mercaptobenzoic acid (4-MBA) molecule ethanol solution is deposited on top of the silver substrate and dilute droplets of Ag / Au nanoparticles are suspended on top of the prepared substrate. The sample is fixed to a piezo-stage for precision positioning and scanning at different positions of the focused beam.
\par 
Fig. \ref{fig4-3}b-d show the experimental radiation pattern at the back focal plane of the objective lens when an Ag-nanoparticle is located at one of the three positions of the radially polarized focused source beam. It is observed that each of the profiles has a donut-shaped structure at the center (corresponding to the transmitted field) and a ring / arc shaped pattern outside (corresponding to SPCE signal). Fig. \ref{fig4-3}e shows the theoretically calculated longitudinal and transverse intensity profiles of the radially polarized light at the focal plane. The point $A$ in Fig. \ref{fig4-3}e marks the geometrical center of the focusing system when longitudinal field component dominates and the transverse component becomes negligible, while at the point $B$, the longitudinal component vanishes. Thus by changing the position of the Ag particle using the piezo-stage, the longitudinal and transverse field components of focused light which interact with the nanoparticle can be changed which in-turn tunes the unidirectional emission of SPP.
\par 
The radiation pattern from the nanoparticle as shown in  Fig. \ref{fig4-3}b appears like that of an electric dipole oscillating in vertical field (longitudinal dipole) and the corresponding SPCE assumes the form of a ring when the Ag-particle is located at the position marked as $A$ in the Fig.\ref{fig4-3}e. This happens because of the axial-symmetry of excitation system in such a situation. When the Ag-particle is placed at the point $B$ of Fig. \ref{fig4-3}e, the longitudinal component vanishes and the transverse component dominates and in this case the radiation pattern is similar to a transverse dipole and the corresponding SPCE gets converted into a bright two-arcs structure (Fig. \ref{fig4-3}d). When the Ag particle is located anywhere between the two-points, transverse and longitudinal components of the fields has a phase difference of $\pi/2$ thus forming a transverse spin state. This transverse spin of the focused field then gets coupled to that of the SPP resulting in an unidirectionality of SPCE pattern (Fig. \ref{fig4-3}c).
\par 
Although we discussed results of coupling of transverse spin of focused beam with that of the SPP using an Ag-nanoparticle on Ag-film, use of  Au-nanoparticle on Ag-film showed different intensity distributions as the resonance conditions are dependent on the material refractive index. Thus the resonance spin-coupling are shown to have useful application for particle identification using focused field \cite{nanoscale2018}.
\par 
As summarized above, there have been initial attempts to detect these illusive entities of the transverse SAM and Belinfante's spin-momentum. However, the practical implementation towards spin-based photonic applications remain to be rigorously investigated. Devising systems for enhancing both the magnitude as well as the spatial extent of these quantities is therefore important for practical applications of these unusual spin and momentum components of light. Some of the possible approaches have been discussed in the preceding sections.

\chapter*{Conclusions}
\addcontentsline{toc}{chapter}{Conclusions}
In this paper, we have introduced the concept of the transverse SAM and Belinfante’s spin-momentum which have been recently discovered. The pertinent feature of structured optical fields that are responsible for the generation of these unusual spin and momentum densities have been illustrated. Specifically, we have given examples of evanescent field, focused vector beams, surface plasmon polaritons, scattering of plane waves and focused vector beams from micro and nanostructures where these quantities are observed. We have also discussed possible ways to enhance both the spatial extent and magnitude of these quantities by exploiting the interference and mutual coupling of different scattering modes. Examples of initial experimental observations have been summarized. The structure of the spin and momentum densities of such systems are fundamentally interesting, specifically because of their analogy with spin-momentum locking and electron topological insulator. They also have considerable promise for developing novel spin based photonic nano-devices for numerous potential applications. It is hoped that the paper will stimulate further research in this exciting field of spin-optics which is still at its infancy.

\bibliographystyle{unsrt}
\bibliography{references.bib}

 \end{document}